%% file: NSUSYDM.tex
\documentclass[11pt,a4paper]{article}
\usepackage{graphicx}
\usepackage{jheppub}
\usepackage{bm}
\usepackage{amsmath}
\usepackage{amssymb}
\usepackage{amscd}
\usepackage{latexsym}
\usepackage{slashed}
\usepackage{color}
\usepackage{graphicx}
\usepackage{hyperref}
\usepackage[normalem]{ulem}

\newcommand{\met}{E_T^{\rm miss}}

\newcommand{\bea}{\begin{eqnarray}}  
\newcommand{\eea}{\end{eqnarray}}  
\newcommand{\charpm}[1]{\tilde\chi^\pm_{#1}}
\newcommand{\neut}[1]{\tilde\chi^0_{#1}}

\title{Uncovering Natural Supersymmetry via the interplay between the LHC and Direct Dark Matter Detection}

\date{\today}
\author[a]{Daniele Barducci,}
\author[b,c]{Alexander Belyaev,}
\author[d,e]{Aoife K. M. Bharucha,}
\author[f]{Werner Porod}
\author[g]{and Veronica Sanz}

\affiliation[a]{LAPTh, Universit\'e Savoie Mont Blanc, CNRS, 9 Chemin de Bellevue, B.P. 110, F-74941 Annecy~le-Vieux, France}
\affiliation[b]{School of Physics and Astronomy, University of Southampton, Highfield, Southampton SO17 1BJ, UK}
\affiliation[c]{Particle Physics Department, Rutherford Appleton Laboratory, Chilton, Didcot, Oxon OX11 0QX, UK}
\affiliation[d]{Physik Department T31, Technische Universit\"at M\"unchen,James-Franck-Stra\ss e~1, D-85748 Garching, Germany}
\affiliation[e]{ CNRS, Aix Marseille U., U. de Toulon, CPT, UMR 7332, F-13288, Marseille, France}
\affiliation[f]{Institut f\"ur Theoretische Physik und Astrophysik, Universit\"at  W\"urzburg, D-97074  W\"urzburg, Germany}
\affiliation[g]{Department of Physics and Astronomy, University of Sussex, Brighton BN1 9QH, UK}

\abstract{
We have explored Natural Supersymmetry (NSUSY) scenarios with low  values of the $\mu$ parameter 
which are characterised by higgsino-like Dark Matter (DM) and compressed spectra for the lightest MSSM particles, $\chi^0_1$, $\chi^0_2$ and $\chi^\pm_1$. This scenario could be probed via monojet signatures, but as the signal-to-background ratio (S/B) is low
we demonstrate that the 8 TeV LHC cannot obtain limits on the DM mass beyond those of LEP2.
On the other hand, we have  found, for the 13 TeV run of the LHC, that by optimising kinematical cuts we can bring the S/B ratio up to the 5(3)\% level
which would allow the exclusion of the DM mass up to 200(250)~GeV respectively, significantly extending LEP2 limits.
Moreover, we have found that LUX/XENON1T and LHC do play very  complementary roles in exploring the parameter space of NSUSY, 
as the LHC has the capability to access regions where DM is quasi-degenerate with other higgsinos, which are challenging for direct detection experiments.
}

\emailAdd{barducci@lapth.cnrs.fr}
\emailAdd{a.belyaev@soton.ac.uk}
\emailAdd{aoife.bharucha@cpt.univ-mrs.fr}
\emailAdd{porod@physik.uni-wuerzburg.de}
\emailAdd{v.sanz@sussex.ac.uk}

\begin{document}

\begin{flushright}
  LAPTH-017/15     \\
TUM-HEP-990/15
\end{flushright}

\maketitle

\tableofcontents

\input{01_Intro.tex}

\input{02_Para-space.tex}

\input{03_DM-DD.tex}
\input{04_LHC.tex}

\input{05_Conclusions.tex}

\newpage
\clearpage
\bibliographystyle{JHEP}
\bibliography{NSUSYDM}
\end{document}

%% file: 01_Intro.tex
\section{Introduction}
The naturalness of Supersymmetry (SUSY), 
which has been subjected to much discussion and thorough investigation for more than two 
decades~\cite{Ellis:1986yg,Barbieri:1987fn,Kane:1993td,Anderson:1994tr,Dimopoulos:1995mi,Chan:1997bi,Akula:2011jx,Liu:2013ula,Chankowski:1997zh,Chankowski:1998xv,Barbieri:1998uv,Kane:1998im,BasteroGil:1999gu,Feng:1999zg,Feng:1999mn,Feng:2012jfa,Feng:2013pwa,Casas:2003jx,Nomura:2005qg,Nomura:2005rj,Kitano:2006gv,Cassel:2009ps,Cassel:2010px,Brust:2011tb,Papucci:2011wy,Hall:2011aa,Blum:2012ii,Espinosa:2012in,D'Agnolo:2012mj,Baer:2012uy,Baer:2012up,Baer:2012cf,Baer:2013bba,Baer:2013jla,Baer:2013vpa,Baer:2014ica,Younkin:2012ui,Fichet:2012sn,Kribs:2013lua,Hardy:2013ywa,Kowalska:2013ica,Kowalska:2014hza,Han:2013kga,Dudas:2013pja,Arvanitaki:2013yja,Martin:2013aha,Boehm:2013qva,Bomark:2013nya,Fan:2014txa,Gherghetta:2014xea,Delgado:2014vha,Delgado:2014kqa,
Fowlie:2014xha,Mustafayev:2014lqa,Antoniadis:2014eta,Casas:2014eca,Baer:2014yta},
has become even more relevant now, as the Large Hadron Collider (LHC) collaborations ATLAS and CMS have started to probe
SUSY in the TeV region.
Indeed, the lack of evidence for superparticles at the CERN LHC, along with the rather high value
of the Higgs boson mass in the context of the  Minimal Supersymmetric Standard Model
(MSSM), raises the question of whether the remaining allowed parameter space 
suffers from a high degree of fine-tuning, and if there is any parameter space of Natural SUSY (NSUSY) left.
We discuss this problem in the framework of the well motivated MSSM.

Based on standard measures of fine tuning~\cite{Ellis:1986yg,Barbieri:1987fn}, the NSUSY parameter space was originally associated with light higgsinos, gluinos and stops, the SUSY partners of the SM Higgs, gluons and top quark.
The present LHC limits on the masses of the latter two are approaching the TeV scale, under the assumption that 
the mass gap between these states and the lightest supersymmetric particle (LSP) , $\neut{1}$, is large enough (see for example ~\cite{Chatrchyan:2013wxa,Chatrchyan:2013lya,Aad:2014pda,Aad:2014wea,Aad:2014kra}). Here $\neut{1}$ is the lightest of the four neutralinos, the mass eigenstates arising from the mixing of the fermionic component of the Higgs and gauge superfields,  $(\tilde H^0_d,\tilde H^0_u)$ and $(\tilde B^0,\tilde W^0)$, which are commonly known as the higgsinos, the bino and the wino.
Moreover, these limits have a certain degree of model dependence and for stops, $\tilde t_{1,2}$, the  experimental limits rely on certain decay channels being dominant
(e.g. $\tilde t_1 \to t \neut{1}$, with $t$ the top-quark) or on a substantial mass splitting between $\tilde t_1$ and $\neut{1}$, and can be significantly relaxed. For example, in the scenario under consideration in this paper with higgsino like dark matter (DM),
the stop branching ratios strongly depend on the left-right admixture of the lightest stop. Therefore {\it model independent} collider bounds on stops are weak or 
non-existent. 

It has however been shown that usual fine-tuning measures, defined as the sensitivity of the weak scale to fractional variations in the fundamental parameters of the theory, can be low even if the masses of the supersymmetric scalars 
are large. 
This happens in the so called ``hyperbolic branch"(HB)~\cite{Chan:1997bi} or ``focus point" 
(FP)~\cite{Feng:1999mn,Feng:1999zg,Feng:2011aa} regions of the minimal super gravity (mSUGRA) parameter space, where the value of the Higgs mass parameter, $\mu$, can be low if the universal gaugino mass $M_{1/2}$ is not too large, as a consequence of the subtle interplay between the electroweak (EW) gauge couplings and the top-Yukawa coupling in the  evolution of the 
squared Higgs mass parameters using the renormalization group equations (RGE).
Moreover it was recently argued~\cite{Baer:2013gva} that EW fine-tuning in SUSY scenarios 
can be grossly overestimated by neglecting additional terms, stemming from the ultra-violet (UV) completion of the model, that can lead to large cancellations favouring a low $\mu$ parameter, but not necessarily a low stop mass up to a certain limit.
Taking this point of view, we will take a low $\mu$ parameter to be the definition of NSUSY throughout our study.
	
In the case $\mu \ll M_1, M_2$ (the EW gaugino mass parameters) one finds that the three lightest 
neutralino and chargino mass eigenstates, $\neut{1}$, $\neut{2}$ and $\charpm{1}$, are quasi-degenerate and
that these states are nearly pure higgsinos. In this  scenario the DM relic density is typically below the
WMAP~\cite{Hinshaw:2012aka} and PLANCK~\cite{Ade:2013zuv} measurements, because of the high rate of higgsino annihilation to standard model (SM) gauge and Higgs bosons and the higgsino coannihilation processes~\cite{Feng:2000zu,Baer:2002fv}.
One should note that this parameter space with the relic abundance {\it below} the experimental constraints is not however excluded, since the remaining relic abundance can be accounted for by other additional sources, e.g. axions.
This NSUSY scenario, which is characterised by relatively light higgsinos in comparison to other SUSY particles,
is not just motivated by its simplicity, but also by the lack of evidence for SUSY to date.
We take advantage of the fact that NSUSY scenarios can be effectively described by a two dimensional parameter space, defined  by
the DM mass, i.e.~the mass of $\neut{1}$, and $\Delta $M, the mass difference between the DM candidate and the next to lightest 
supersymmetric  (NLSP), typically $\charpm{1}$, and our study explores the complementarity of the LHC and direct detection (DD) DM search experiments in covering this region.
Such complementarity was the subject of recent studies, see e.g.\cite{Crivellin:2015bva} and references there in.

It was already shown a decade ago that the HB/FP parameter space is challenging to probe at the LHC~\cite{Baer:2004qq} 
even if the mass gap between $\neut{1,2}$ and $\charpm{1}$ is large enough to provide leptonic signatures. The most challenging case arises when the mass gap between these states is too small to produce any detectable leptons.
The  only way to probe such a scenario is via mono-object signatures, i.e. signatures involving a high transverse momentum particle recoiling against missing transverse emerge ($\met$), of which the monojet signature is particularly of relevance at the LHC, as initially suggested in~\cite{Alves:2011sq} for generic compressed spectra.

This technique has already been used in studies of quasi-degenerate higgsino
spectra via monojet+$\met$ and monojet+$\met$+soft di-lepton signatures for the NSUSY parameter space we consider~\cite{Han:2013usa,Schwaller:2013baa,Han:2014kaa,Baer:2014cua,Baer:2014kya}.
However, we believe that these analyses are not entirely complete and/or have certain drawbacks.
For example, in~\cite{Han:2013usa} the 95\% confidence level (CL) reach for the 14 TeV run of the LHC was calculated assuming a signal (S) to background (B) ratio below the 2\% level, which is probably not quite feasible when taking into account that the
actual systematic error should be above 3-5\% even for quite optimistic analyses~\cite{Khachatryan:2014rra,ATLAS:2012zim,ATL-PHYS-PUB-2014-007}. In~\cite{Han:2014kaa} the authors have performed their analysis at the parton level while, as shown in a preliminary analysis
\cite{Brooijmans:2014eja,SUSY-Belyaev}, a fast detector simulation analysis leads to qualitatively different results and, therefore, is crucial.
In~\cite{Baer:2014cua} the conclusion about the observability of the quasi-degenerate higgsino NSUSY scenario from the monojet search was negative. However in this study the authors did not attempt to optimise the $\met$
which turns out to be important as we will show in this paper. In ~\cite{Schwaller:2013baa} the prospects
were more optimistic even after including systematics uncertainties. However, also there  no  optimisation of
the cuts was performed which considerably enhances the accessible mass range as we will show. 
One should also note the Ref.~\cite{Baer:2014kya} where authors suggested a new promising signature including a pair of soft leptons, and have demonstrated its potential power. However, in this study the important $b\bar{b}$ background was not considered as we discuss below, implying thus further background investigation for this signature which we do not consider at present.
Finally one should mention Ref.~\cite{Han:2015lma}, which studied similar to~\cite{Baer:2014kya} monojet plus soft lepton signature suggesting visibly harder cuts to suppress $b\bar{b}$ background. The respective higgsino mass reach from this study is quite limited. On the other hand  the suggested b-jet veto will not quite work for the signature with isolated soft muons,
so, we believe that one should estimate $b\bar{b}$ background more precisely even for the case of harder cuts suggested in~\cite{Han:2015lma}.

Motivated by the above-mentioned previous studies, here we aim to perform a comprehensive and realistic analysis of the monojet potential to probe this NSUSY scenario.
Our analysis is performed at the level of a fast detector simulation and the whole two-dimensional NSUSY parameter space mentioned above, rather than selected benchmarks as were attempted previously, is explored therefore completely covering the region of our interest.
We consider prudent systematic errors and optimise the kinematic cuts to 
keep the S/B ratio at a reasonable level. We then discuss the LHC potential to cover the NSUSY parameter space at 8 TeV and produced projections for the 13 TeV run of the CERN machine. By analysing also the exclusion potentiality for DM direct detection experiments, we aim to show that 
collider and DD experiments have a high degree of complementarity.

The paper is organised as follows.
In Section 2 we describe the parameter space and mass spectrum of NSUSY while in Section 3 we discuss the DM properties
of this scenario. Section 4 is dedicated to the analysis of the collider phenomenology of the compressed higgsino scenario while in Section 5 we show the complementarity of collider and the DD experiment. We conclude in Section 6.

%% file: 02_Para-space.tex

\section{Parameter space and Spectrum of NSUSY}

In the bases $(\tilde B^0,\tilde W^0, \tilde H^0_d,\tilde H^0_u)$ and $(\tilde W^0, \tilde H^0_d)$ the mass matrices of the neutralino and chargino sector of the MSSM are
\begin{equation}
M_{\neut{0}}=
\left(
\begin{array}{c c c c}
M_1                      & 0                      & -M_Z s_\omega c_\beta & M_Z s_\omega s_\beta \\
0                        & M_2                    &  M_Z c_\omega c_\beta & -M_Z c_\omega s_\beta \\
-M_Z s_\omega c_\beta    & M_Z c_\omega c_\beta   &                       & -\mu                  \\
M_Z s_\omega s_\beta     &-M_Z c_\omega s_\beta   & -\mu                  & 0                  \\
\end{array}
\right)
\quad\,
M_{\charpm{1}}=
\left(
\begin{array}{c c}
M_2                   & \sqrt{2} M_W s_\beta \\
\sqrt{2} M_W c_\beta  & \mu  \\
\end{array}
\right)
\end{equation}
where $M_1$ and $M_2$ are the soft susy breaking mass parameter for $\tilde B$ and $\tilde W$, $\mu$ is the Higgsino mass parameter, $c_\omega$ and $s_\omega$ are $\cos$ and $\sin$ of the Weinberg angle, $\tan\beta=v_2/v_1$ is the ratio of the vacuum expectation
values of two Higgs doublets, $s_\beta$, $c_\beta$ are $\sin \beta$ and $\cos \beta$, respectively and $m_Z$, $m_W$ are the masses of the SM gauge bosons $Z^0$ and $W^\pm$

As a first step we consider scenarios where the $\neut{1}$ and $\charpm{1}$
have a high Higgsino component with all the SUSY partners of the SM fermions  with masses in the multi TeV range and, 
to first obtain a qualitative understanding of the spectrum, we then expand the corresponding mass eigenvalues in 
 the limit $|\mu|\ll|M_1|,|M_2|$ obtaining:
\begin{eqnarray}
m_{ \neut{1,2} } &\simeq& \mp \left[ |\mu|  \mp \frac{m_Z^2}{2} (1\pm s_{2\beta}) \left(\frac{s_\omega^2}{M_1} +\frac{c_\omega^2}{M_2}\right)\right]\\
m_{\charpm{1} } &\simeq& 
 |\mu| \left(1+\frac{\alpha(m_Z)}{\pi}\left(2+\ln\frac{m^2_Z}{\mu^2}\right)\right)
- s_{2\beta}  \frac{m^2_W}{M_2} 
\end{eqnarray}
 where we have defined $s_{2\beta} = \sin(2 \beta) sgn(\mu)$ and $\alpha$ is the electromagnetic structure constant.
In the case of $\charpm{1}$ we have also included the electromagnetic corrections as this shifts
the mass by about 0.5~\% which is indeed important in those cases where the mass splitting between $\charpm{1}/\neut{2}$ and $\neut{1}$ is of the order of a few GeV.
We would like to note, that in our numerical results we
have included the complete one-loop corrections in the calculation of the masses as this changes the absolute
mass scale by several per-cent. However, the mass splittings discussed below are hardly affected as
the additional corrections changes those at most by $O(\alpha /(4 \pi) \ln(m^2_W/m^2_Z)$ which is below
the per-mile level. 
For positive (negative) $\mu$ the mass eigenstate with negative
(positive) CP-eigenvalue is the lightest one. 
The mass splittings are given by
\begin{eqnarray}
\Delta m_0 & = &  m_{\neut{2}} -  m_{\neut{1}} \simeq m_Z^2 \left(\frac{s_\omega^2}{M_1} +\frac{c_\omega^2}{M_2}\right) \\
\Delta m_\pm & = & m_{\charpm{1}} -  m_{\neut{1}} \simeq \frac{\Delta m_0}{2}  + |\mu| \frac{\alpha(m_Z)}{\pi}\left(2+\ln\frac{m^2_Z}{\mu^2}\right)
\end{eqnarray}
where we have neglected corrections of the order $1/\tan\beta$ and $(\mu/M_{1,2})^2$.

In order to analyse scenarios where the mass splitting between $\charpm{1}$ and $\neut{1}$
varies from the (quasi) degenerate regime up to the regime with larger mass splittings, we have chosen
the following range of parameter space:
\begin{equation}
\mu=(100, 300) \textrm{ GeV} \quad M_1= (\mu, \mu+600) \textrm{ GeV and } (-\mu,-\mu-600) \textrm{ GeV} \quad \tan{\beta}= (5-50),
\label{eq:param_space}
\end{equation}
fixing the value of $M_2=$ 2 TeV, which has the effect of decoupling $\chi^0_4$ and $\chi^\pm_2$ that will not be considered anymore in the following, along with the rest of SUSY spectrum, which is assumed to be decoupled.

For $|M_1| \simeq \mu$  one can obtain a simple approximation for  $\neut{1,2,3}$ masses~\cite{ArkaniHamed:2006mb} which we confirm by numerical evaluation, indicating that in this parameter region $\neut{1}$ and $\neut{3}$ are strongly mixed bino-higgsino states whereas $\neut{2}$ is essentially a higgsino-like state with only a small bino component. 

\begin{figure}[htb]
\includegraphics[width=0.48\textwidth]{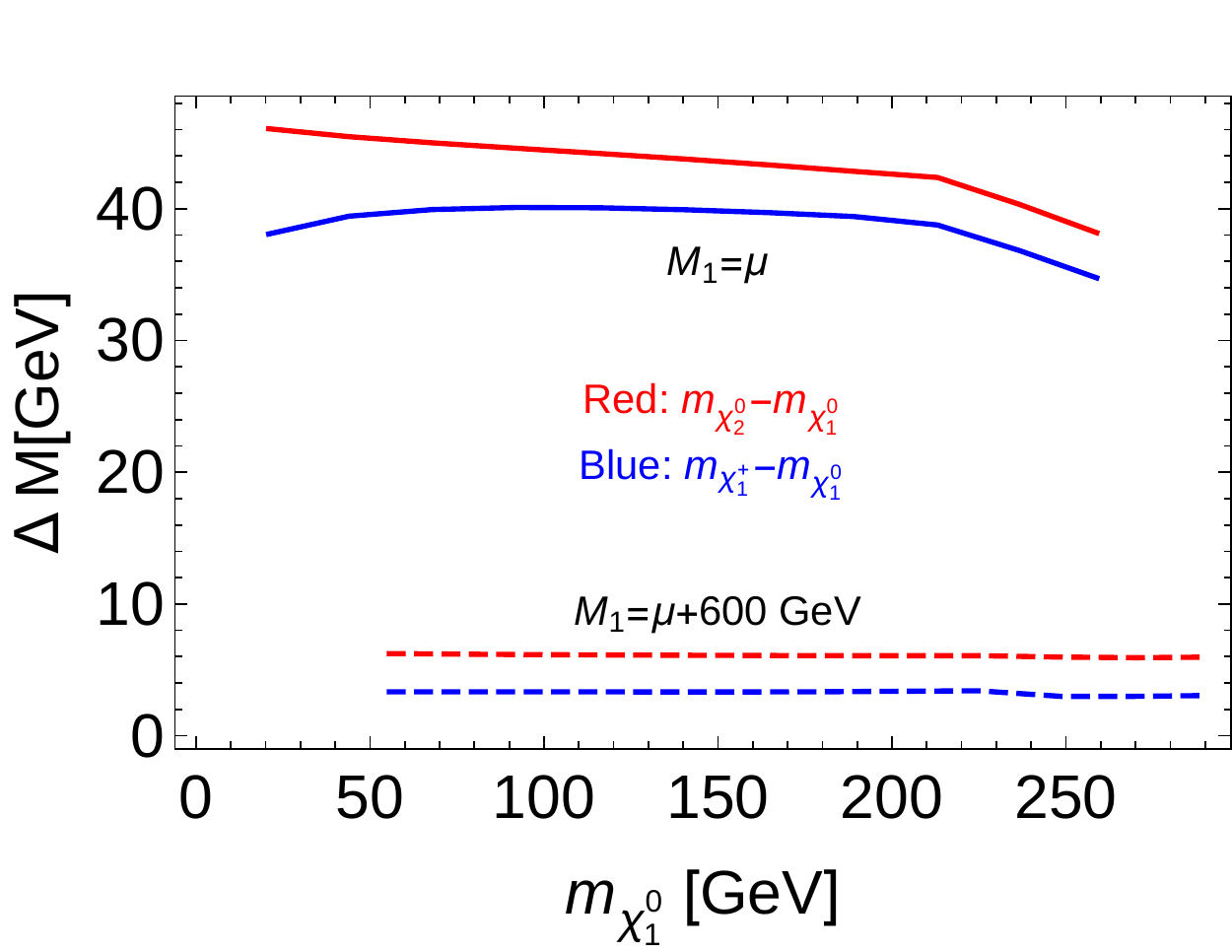}{}\hfill
\includegraphics[width=0.48\textwidth]{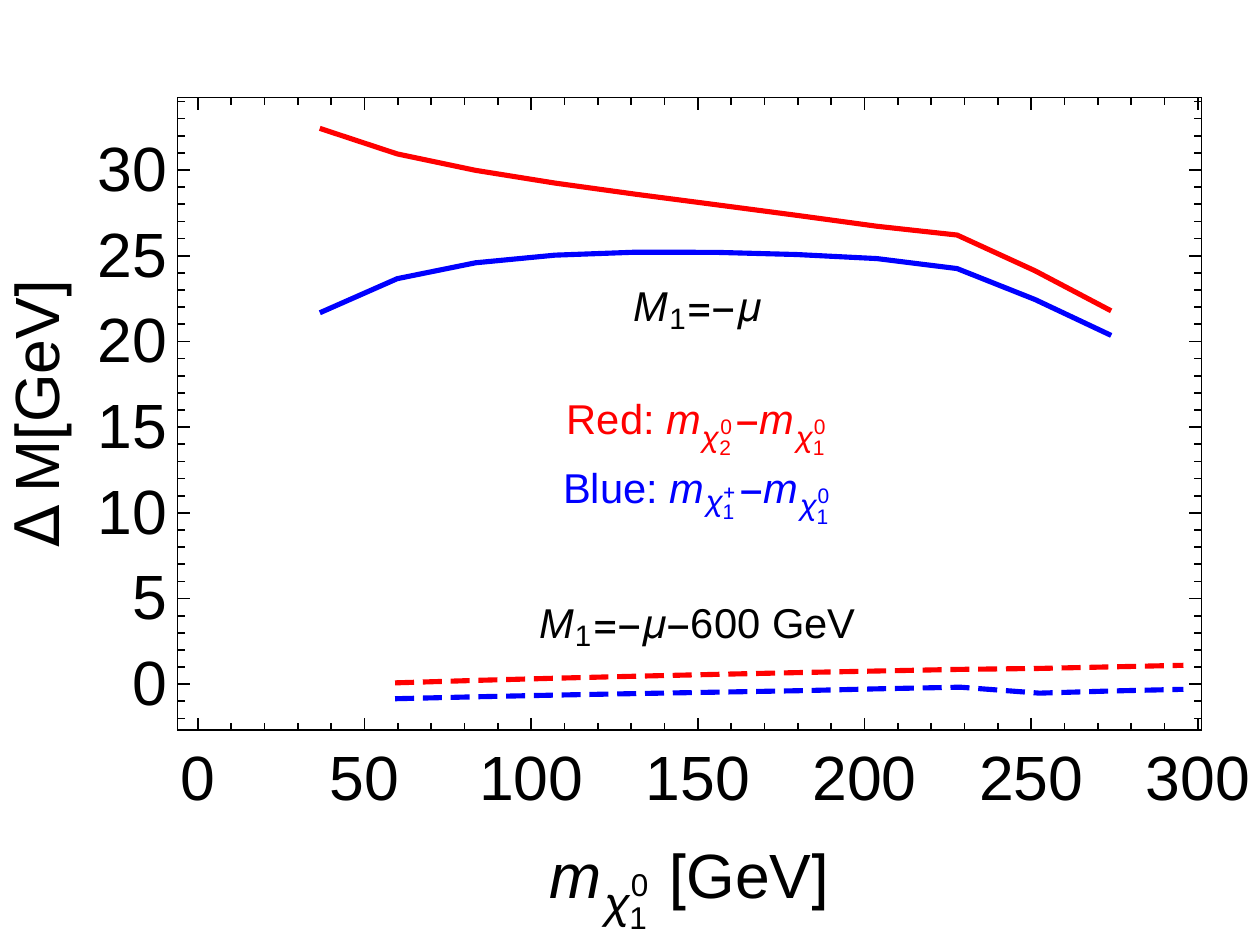}{}\\
\includegraphics[width=0.48\textwidth]{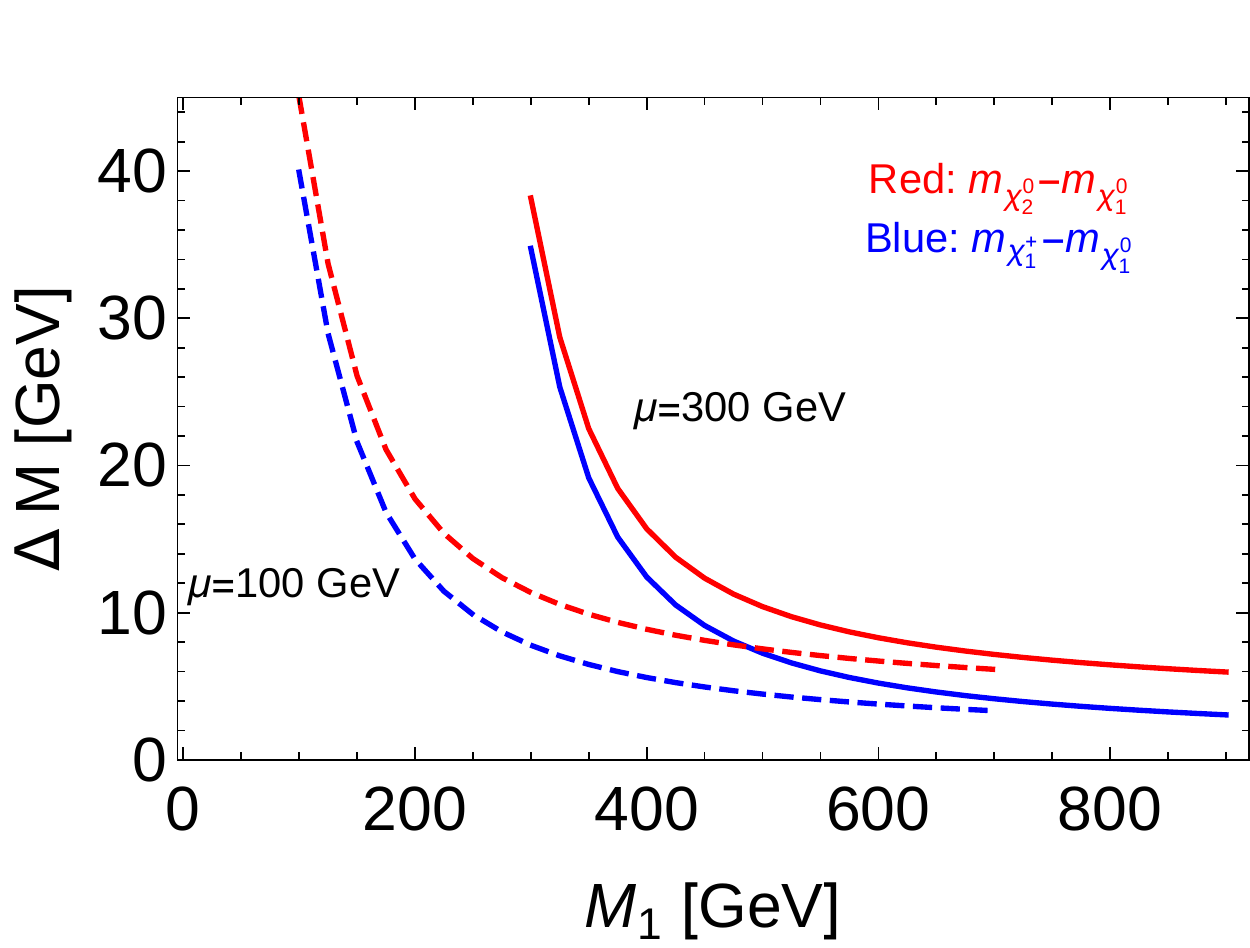}{}\hfill
\includegraphics[width=0.48\textwidth]{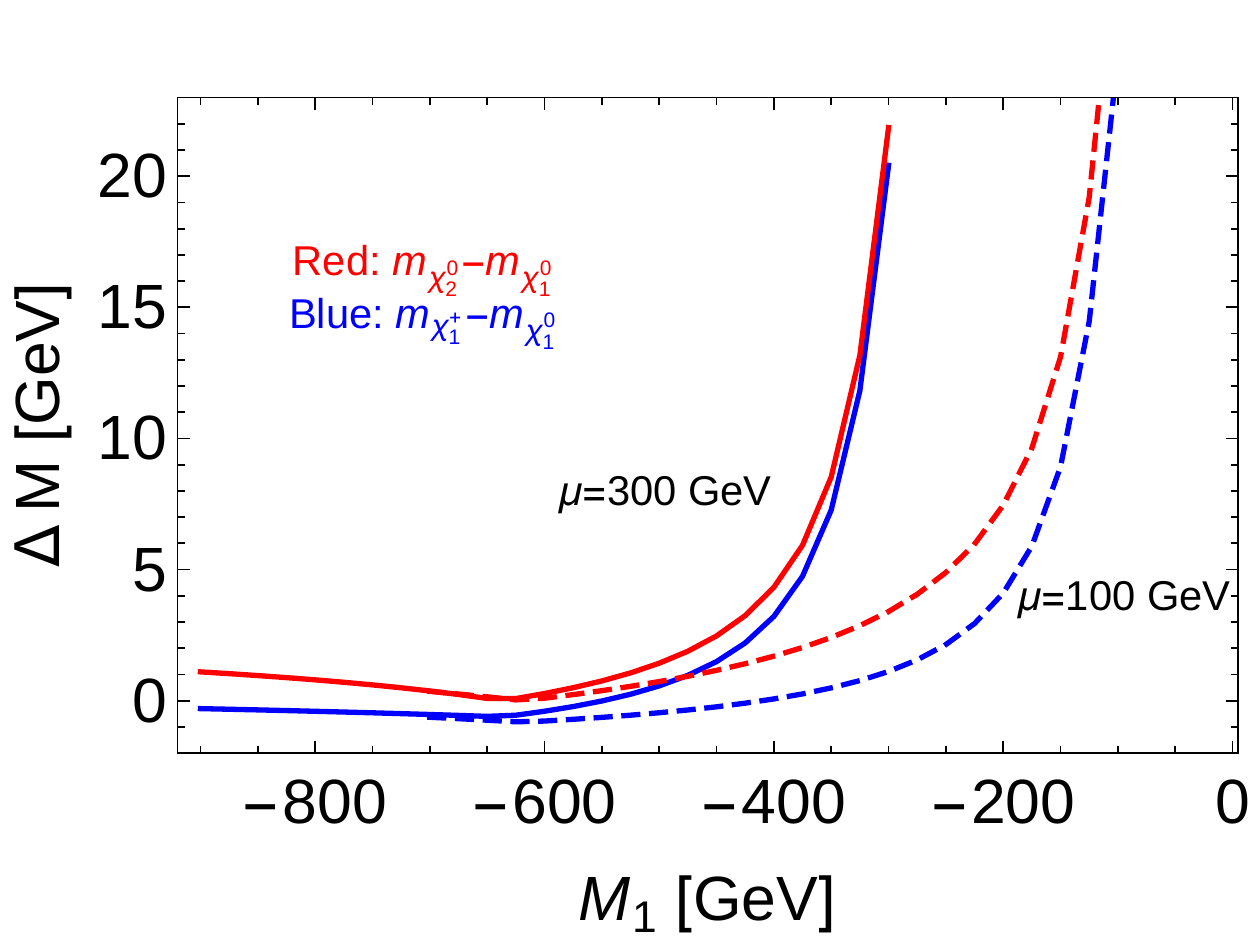}{}\\
\caption{$\charpm{1}-\neut{1}$ and $\neut{2}-\neut{1}$ mass splitting values as a  function of $m_{\chi^0_1}$ (upper row)
and $M_1$ (lower row) for the case $M_1>0$ (left) and $M_1<0$ (right).}
\label{fig:dm-mu-m1}
\end{figure}

\begin{figure}[h!]
\includegraphics[width=0.43\textwidth]{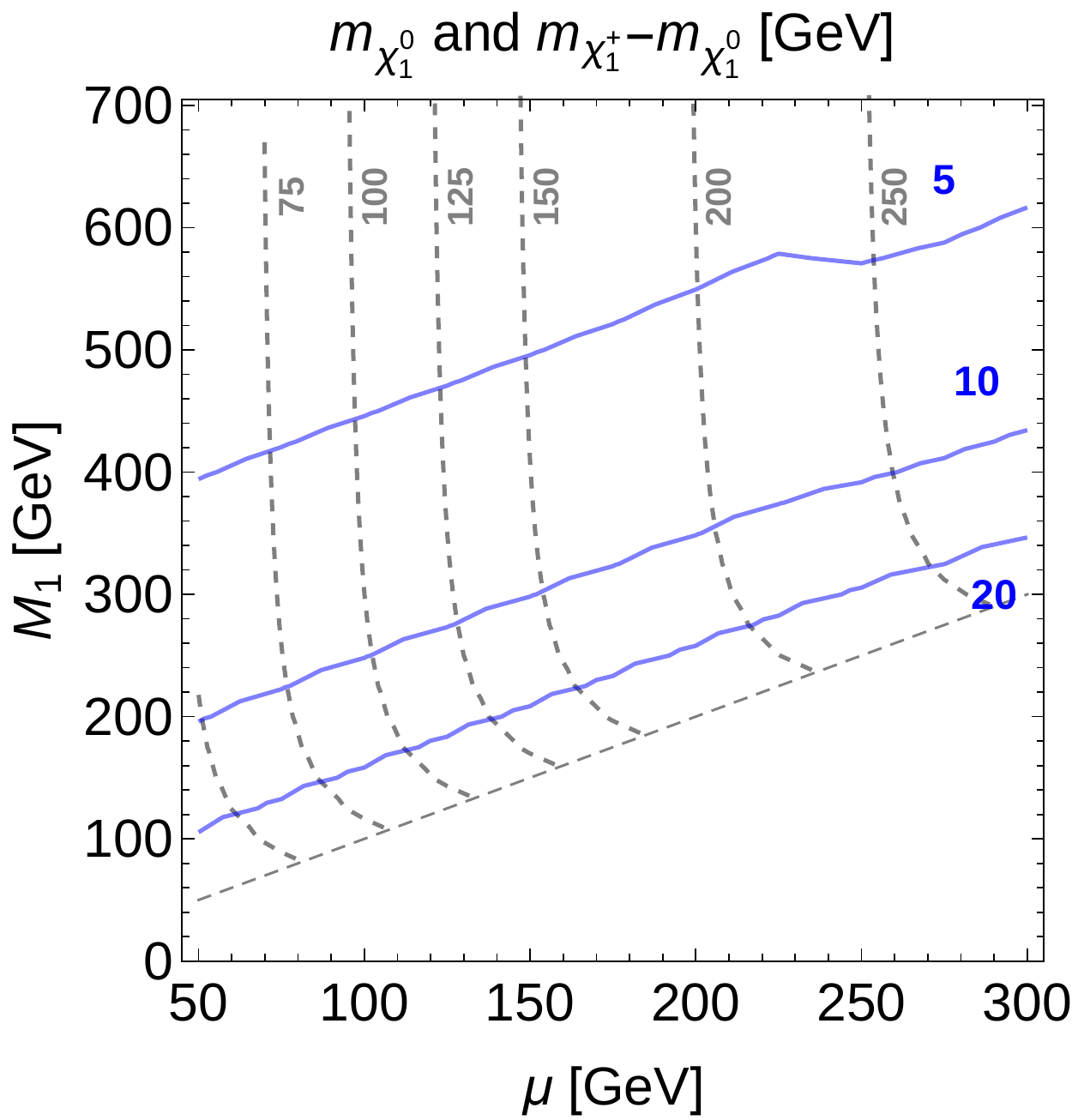}\hfill
\includegraphics[width=0.45\textwidth]{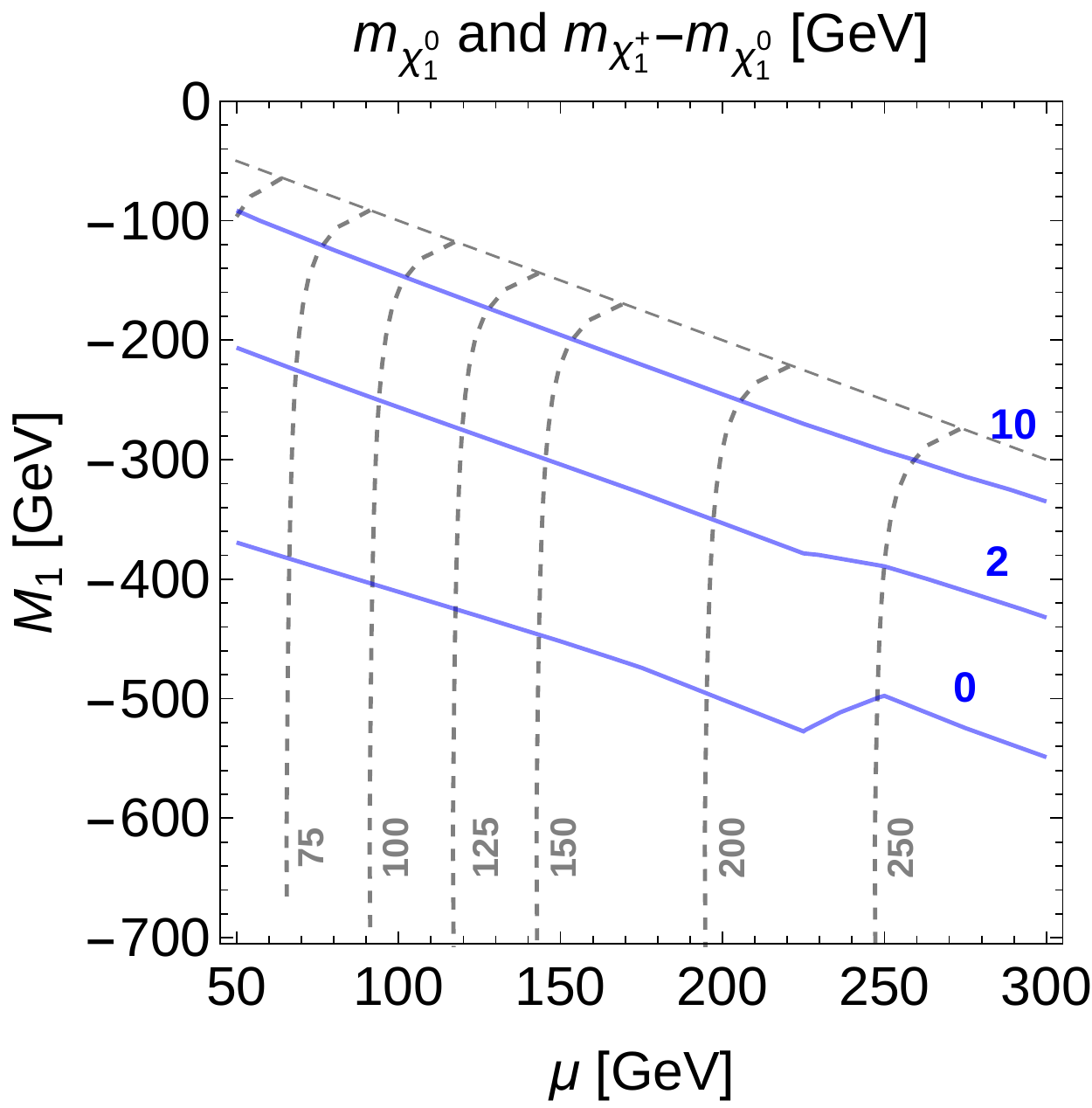}
\caption{Mass splitting between $\charpm{1}$ and $\neut{1}$  (blue solid line) and 
mass of $\neut{1}$ (gray dashed line) in the $\mu$-$M_1$ plane.}
\label{fig:mdm-mu-m1}
\end{figure}

The mass splitting $\Delta$M$=m_{\charpm{1}}-m_{\neut{1}}$ is shown in Fig.~\ref{fig:dm-mu-m1} as a function of $\mu$ and $M_1$ for the case of positive and negative $M_1$ while in Fig.~\ref{fig:mdm-mu-m1} we show the contour lines for $\charpm{1}-\neut{1}$ mass splitting.
The relation between $|M_1|-\mu$ and the value of the mass splitting, which 
runs from quasi-degenerate scenario, $\Delta$M$ \simeq 1-5$~GeV for large $M_1$,
to bigger values, $\Delta$M$ \simeq 10-30$~GeV for $|M_1|\simeq \mu$, is clearly shown in this plots where the mass of the lightest neutralino, the DM candidate, is also presented. 

The nature of the neutralinos and charginos, as well as the small mass splitting between them, has a strong
impact on their decay modes. 
In the case of pure higgsinos, the three body decays are dominated by 
virtual vector bosons, and due to the small mass differences
the decays into third generation fermions are suppressed. 
Note that in the scenario where  $M_1$ is close to
$|\mu|$, the off-shell lightest Higgs boson, $h^0$, can also give sizeable
contributions \cite{Bartl:1999iw,Djouadi:2000aq}.
Also, one should note that in the case $|M_1| \simeq \mu$,
$m_{\neut{3}} - m_{\neut{1}} \simeq 2 (m_{\neut{2}} - m_{\neut{1}})$
and $\neut{3}$ decays to the lightest chargino with the 50\% probability
while sharing  about 25\% decay to  each of   $\neut{1}$ and  $\neut{2}$.

 Three body decays in the limit of small mass separation are discussed in~\cite{DeSimone:2010tf}, where an 
effective theory study of the pseudo-Dirac DM scenario~\cite{Nelson:2002ca,Hsieh:2007wq,Belanger:2009wf} such as the higgsino-like was performed. In this limit the decay width does not depend on the overall neutralino mass, but just on the mass difference
 \begin{eqnarray}
 \Gamma( \tilde{\chi}^\pm_{1}, \tilde{\chi}^0_{2} \to f\, f' \,  
 \tilde{\chi}^0_{1}) = \frac{C^4}{128 \pi^3} \frac{\Delta m^5}{\Lambda^4}
 \end{eqnarray}
where $\Lambda \simeq m_{W,Z,h^0}$ is the mass of the leading mediator, while $\Delta m$ is either
$m_{\neut{2}}-m_{\neut{1}}$ or 
$m_{\charpm{1}}-m_{\neut{1}}$. 
As an example for off-shell $Z$ exchange and decay into leptons, the coefficient $C$ is 
\begin{eqnarray}
C^4 \simeq 
  \frac{1}{4}  \frac{g^4}{c^4_\omega} \left( (s^2_\omega-1/2)^2 + s^4_\omega\right)
\end{eqnarray}
and a similar expression occurs for the case of the off-shell $W$-decay.
The proper decay length is very sensitive to the value of $\Delta m$, and values below the GeV 
lead to displaced vertices, or collider-stable $\charpm{1}$ and $\neut{2}$.
Indeed, for the decay 
$\tilde{\chi}^0_{2} \to f\, \bar{f} \,  \tilde{\chi}^0_{1}$ with an off shell $Z$ exchange, the proper decay length is given by
\begin{eqnarray}
L = c \tau \simeq 0.025\textrm{ cm } \left(\frac{\Delta m}{1 \textrm{ GeV}}\right)^{-5} 
\end{eqnarray}
which implies that for $\Delta m \lesssim 0.1$ GeV, $\neut{2}$ would be collider stable. Similarly, for 
$\Delta m \lesssim 1$~GeV one could look for displaced vertices of order 100 $\mu$m. Note that the measured decay length 
would depend on the boost factor of the decaying neutralino to be taken into account as discussed in details in Ref.~\cite{DeSimone:2010tf}.
While collider stable  $\neut{2}$ will contribute to the $\met$,
long lived or collider stable $\charpm{1}$ will provide a clear signature in the detector.
A $\charpm{1}$ with a long enough lifetime can  be detected in the tracking
detectors by identifying decays that result in tracks with no associated hits in the outer region of the
tracking system as recently analysed by ATLAS~\cite{Aad:2013yna} and CM~\cite{CMS:2014gxa} collaborations.
Both collaborations have obtained similar results, concluding on sensitivity for charginos with a lifetime between 0.1 ns and 100 ns and covering chargino mass up to 500 GeV  which significantly
surpass  the reach of the LEP experiments. 
For  $\Delta$M$ \lesssim 0.25$~GeV  the chargino could be a collider-stable charged particle~\cite{DeSimone:2009ws}, and bounds on such a situation arising from the 8 TeV run of the LHC can be estimated to be $m_{\tilde{\chi}^\pm} \gtrsim $ 300 GeV~\cite{ATLAS:2014fka}. 
On the other hand, we have found that for $\Delta$M$ \gtrsim 0.4$~GeV  there is no limit on  $m_{\charpm{1}}> 100$~GeV from the above LHC searches. Therefore, our task is to analyse the potential of the monojet search to cover the NSUSY  parameter space  with  $\Delta$M$ \gtrsim 0.4$~GeV.

%% file: 03_DM-DD.tex
\section{Dark Matter direct and indirect detection in the NSUSY parameter space }

The results from Planck~\cite{Ade:2013zuv,Planck:2015xua} (see also WMAP~\cite{Hinshaw:2012aka}) have further decreased the error on the already very precise measurement of the dark matter relic density, $\Omega_{\rm DM}^{\rm Planck} h^2=0.1184\pm0.0012$. 

\begin{figure}[htb]
\includegraphics[width=0.49\textwidth]{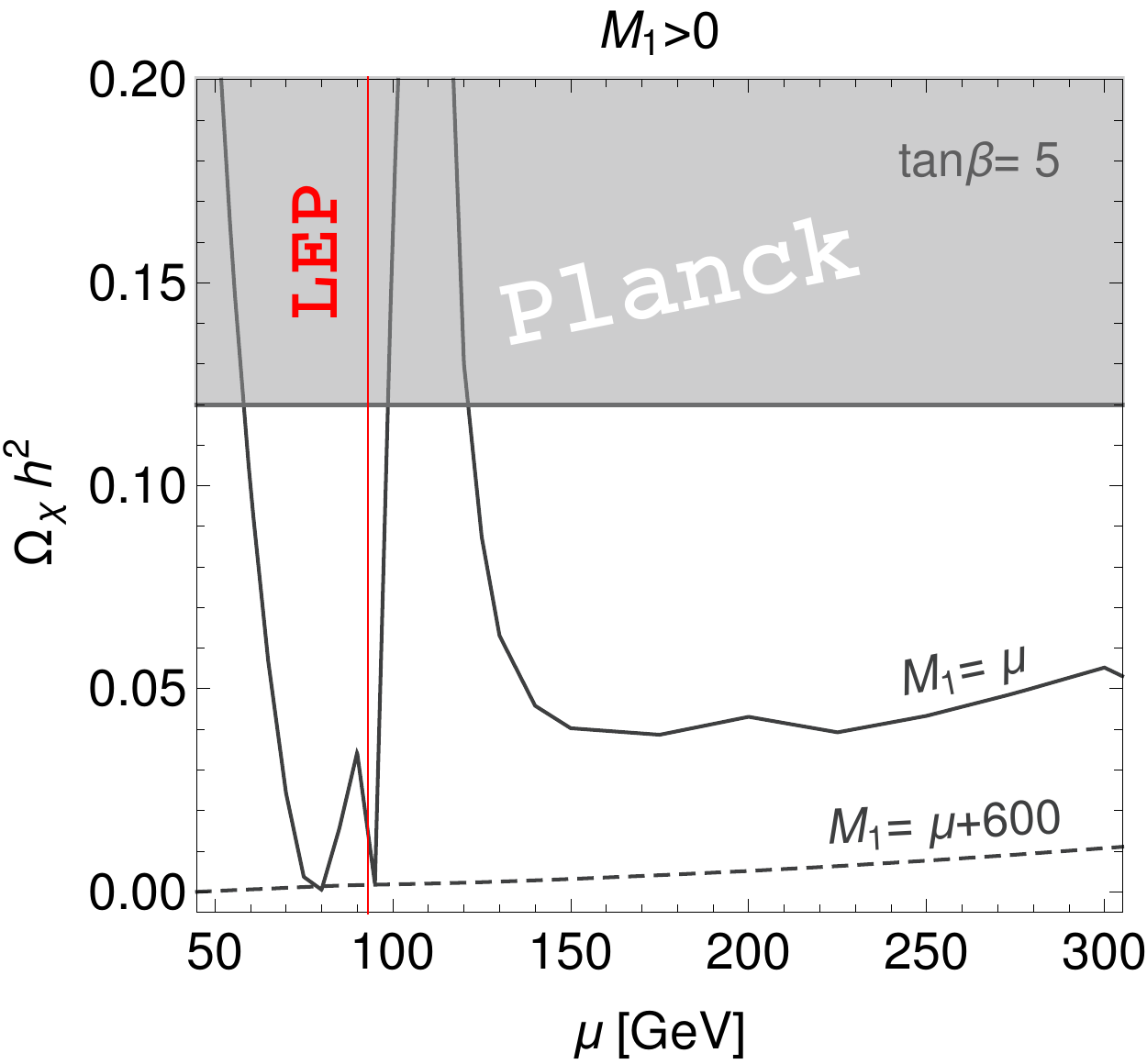}\hspace{.5cm}\includegraphics[width=0.49\textwidth]{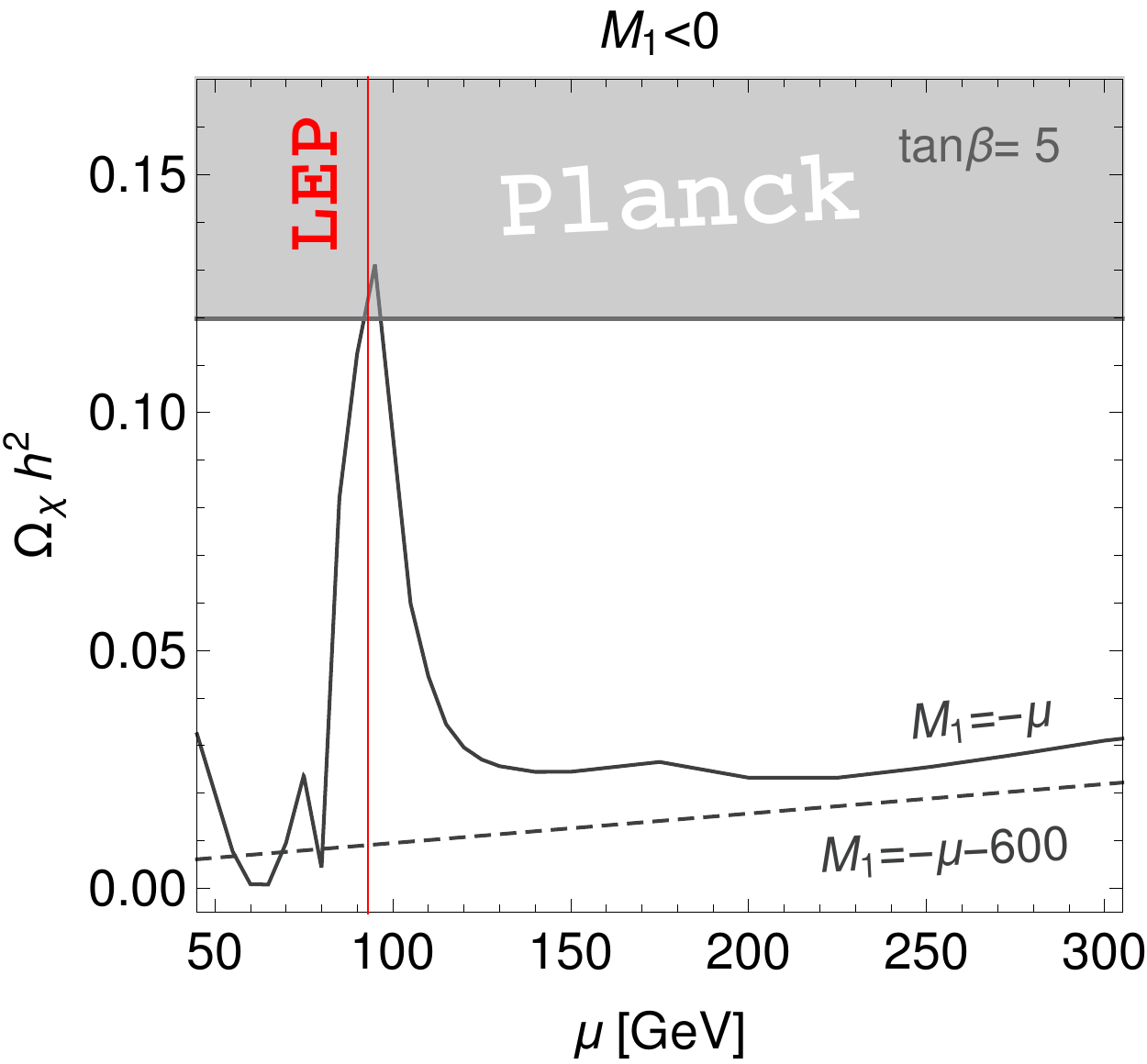}\\
\includegraphics[width=0.49\textwidth]{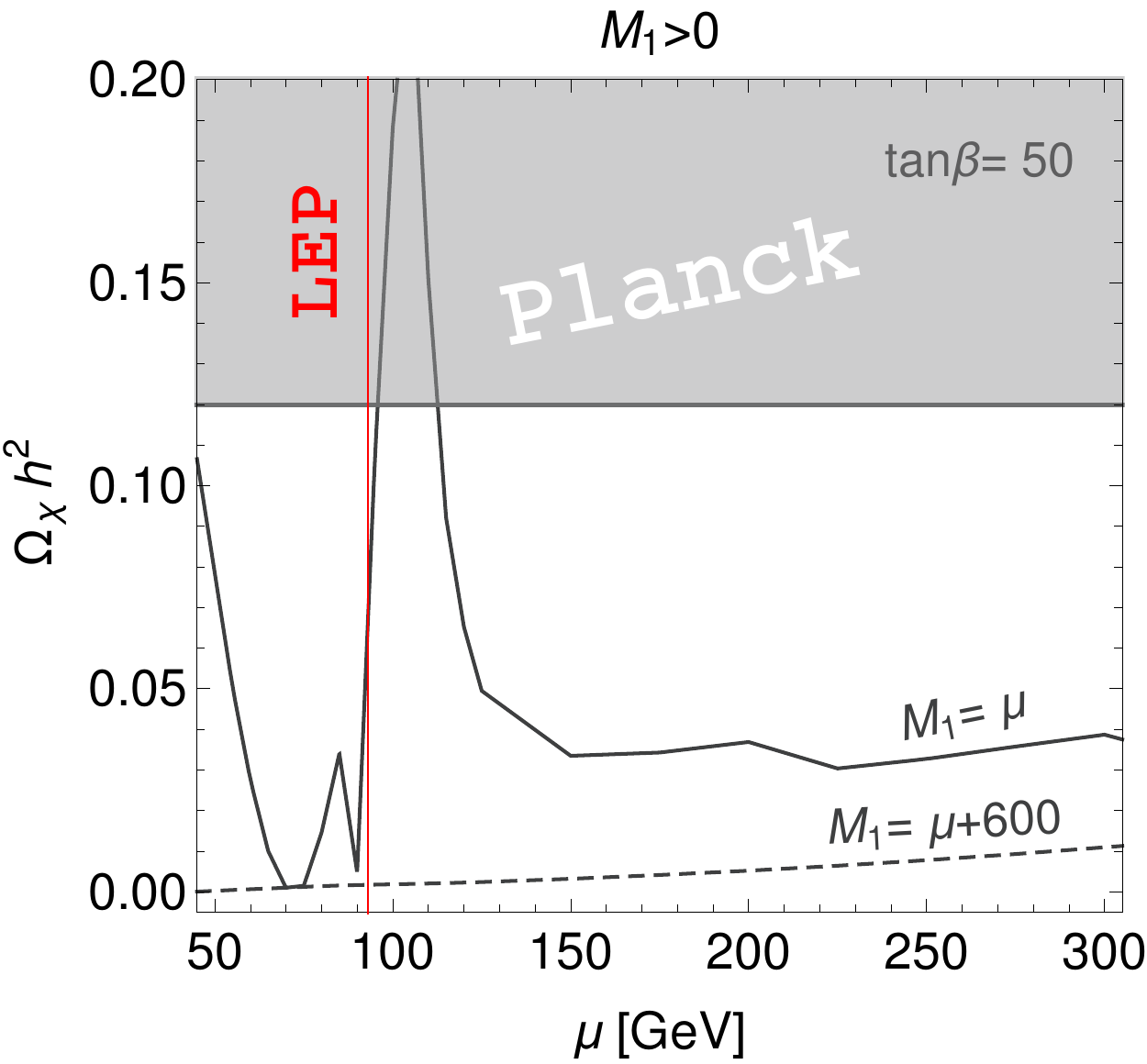}\hspace{.5cm}\includegraphics[width=0.49\textwidth]{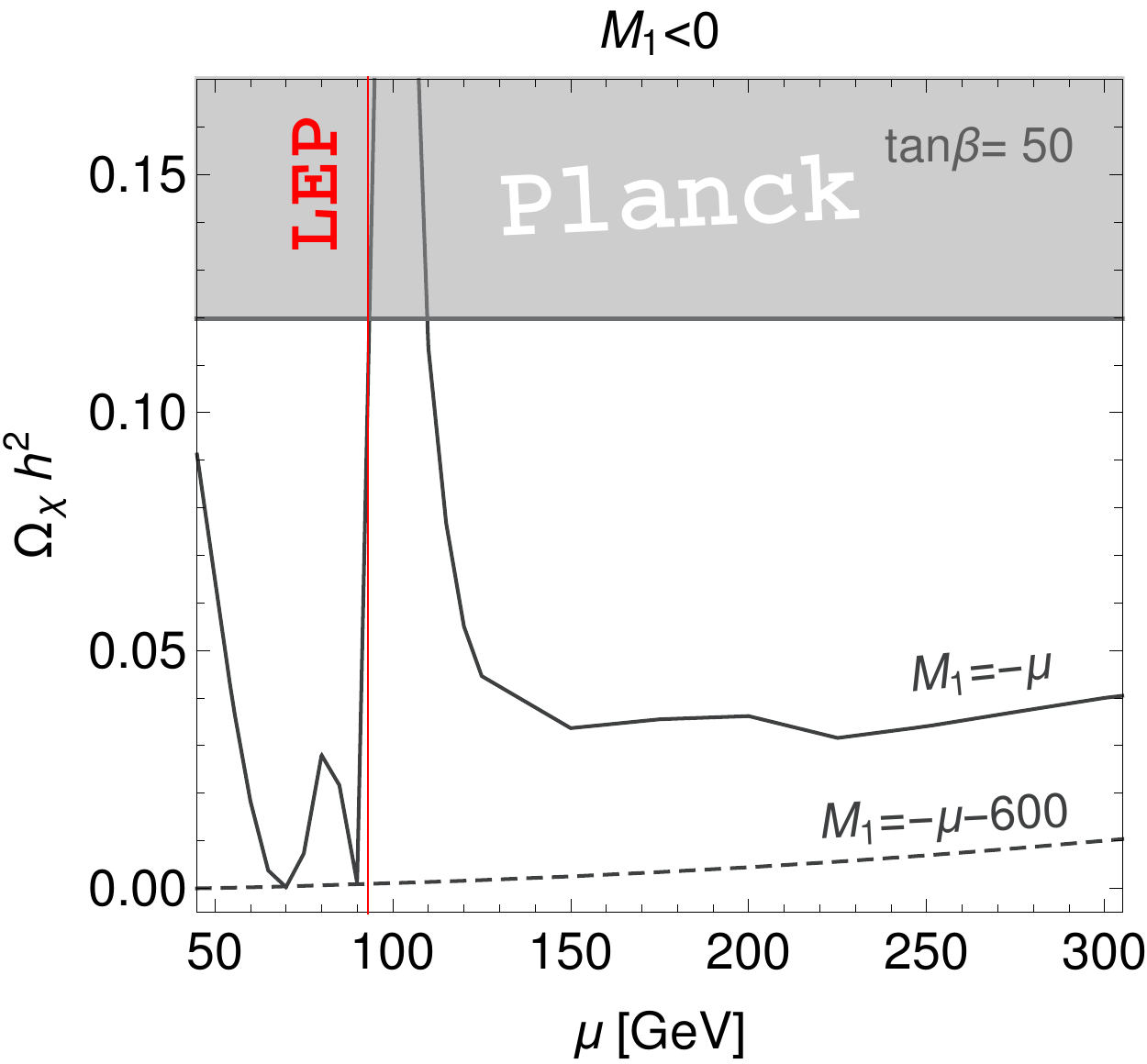}
\caption{The predicted value of the dark matter relic density $\Omega_{\rm DM} h^2$ is shown as a function of $\mu$ for $\tan\beta=5,50$ and positive (left) or negative (right) values of $M_1$ as indicated. The relic density measured by the Planck satellite, $\Omega_{\rm DM}^{\rm Planck} h^2$, is also shown for comparison, and the region excluded due to an overabundance of DM is indicated in grey.} 
\label{fig:Omega}
\end{figure}

As we assume R-parity to hold, the LSP will be stable and will contribute to this relic density.
In the scenarios under consideration, the LSP is the lightest neutralino, $\neut{1}$, which is dominantly higgsino-like with a variable bino component.
It is well known that for $\neut{1}$ of mass $\mathcal{O}(100$ GeV), a higgsino-like LSP 
provides DM relic density below the level  observed by Planck.
This happens because of the annihilation and co-annihilation rate of 
LSP and NLSP particles in the early Universe begin too large.
On the other hand, for bino-like neutralinos the annihilation is suppressed, resulting 
in DM over-abundance. In the mass range we study, mixed bino-higgsino LSPs can therefore lead to the correct relic abundance. For $\mu\lesssim M_1$ the LSP is mainly higgsino-like and the value of $\Omega_{\rm DM}h^2\lesssim\Omega_{\rm DM}^{\rm Planck} h^2$, however this nonetheless at least solves the typical problem of the over-closure of the universe for neutralinos of this mass range in generic SUSY parameter space. In this case we then assume that the remaining relic abundance is accounted for by other means, for example,
it could come from
multi-TeV moduli field where the higgsino LSP is non-thermally produced (see e.g.~Ref.~\cite{Allahverdi:2012wb})
or from  mixed axion-higgsino DM (e.g.~Ref.~\cite{Baer:2012uy}).

\begin{figure}[t]
\includegraphics[width=0.49\textwidth]{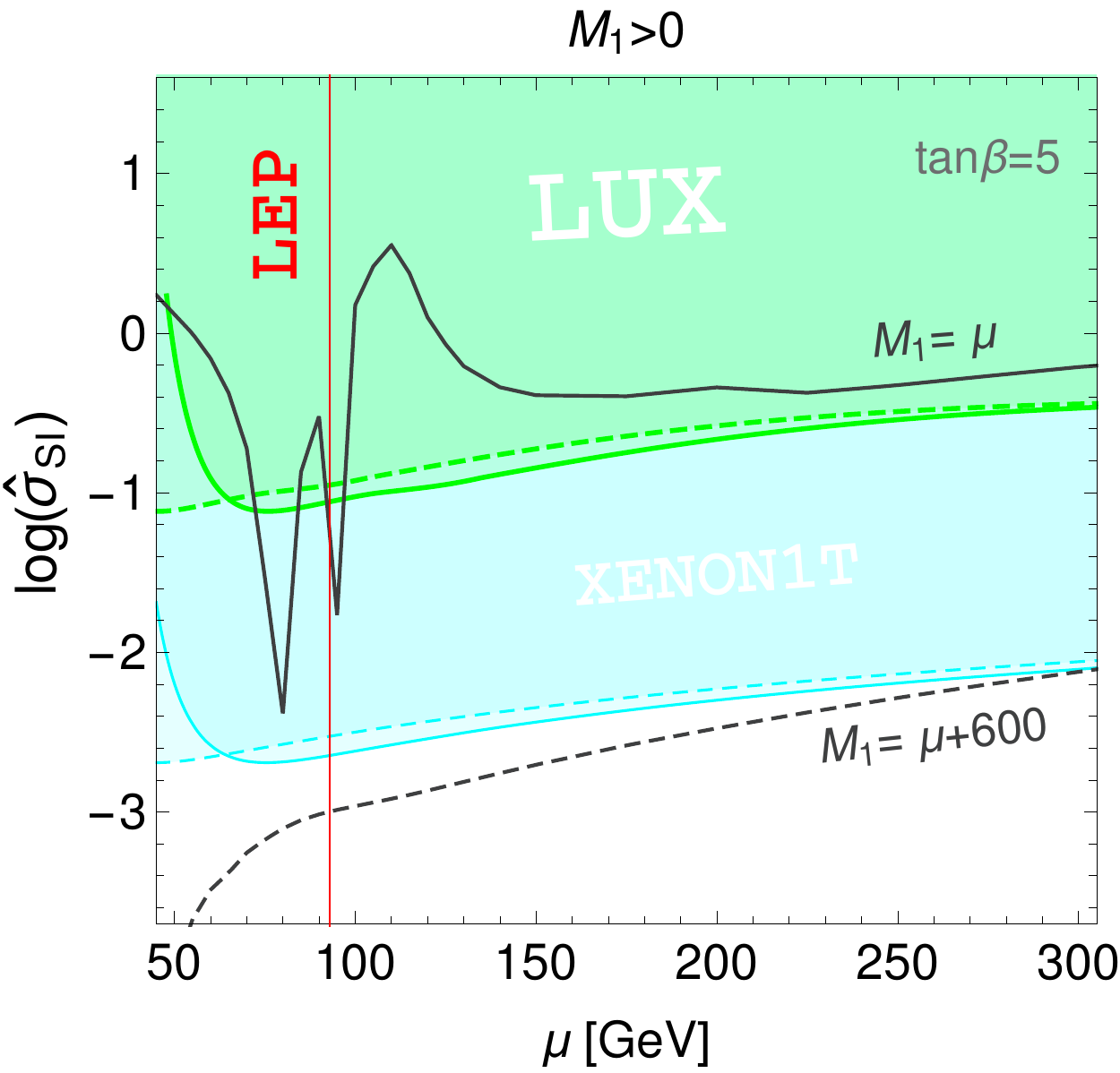}\hspace{.5cm}\includegraphics[width=0.49\textwidth]{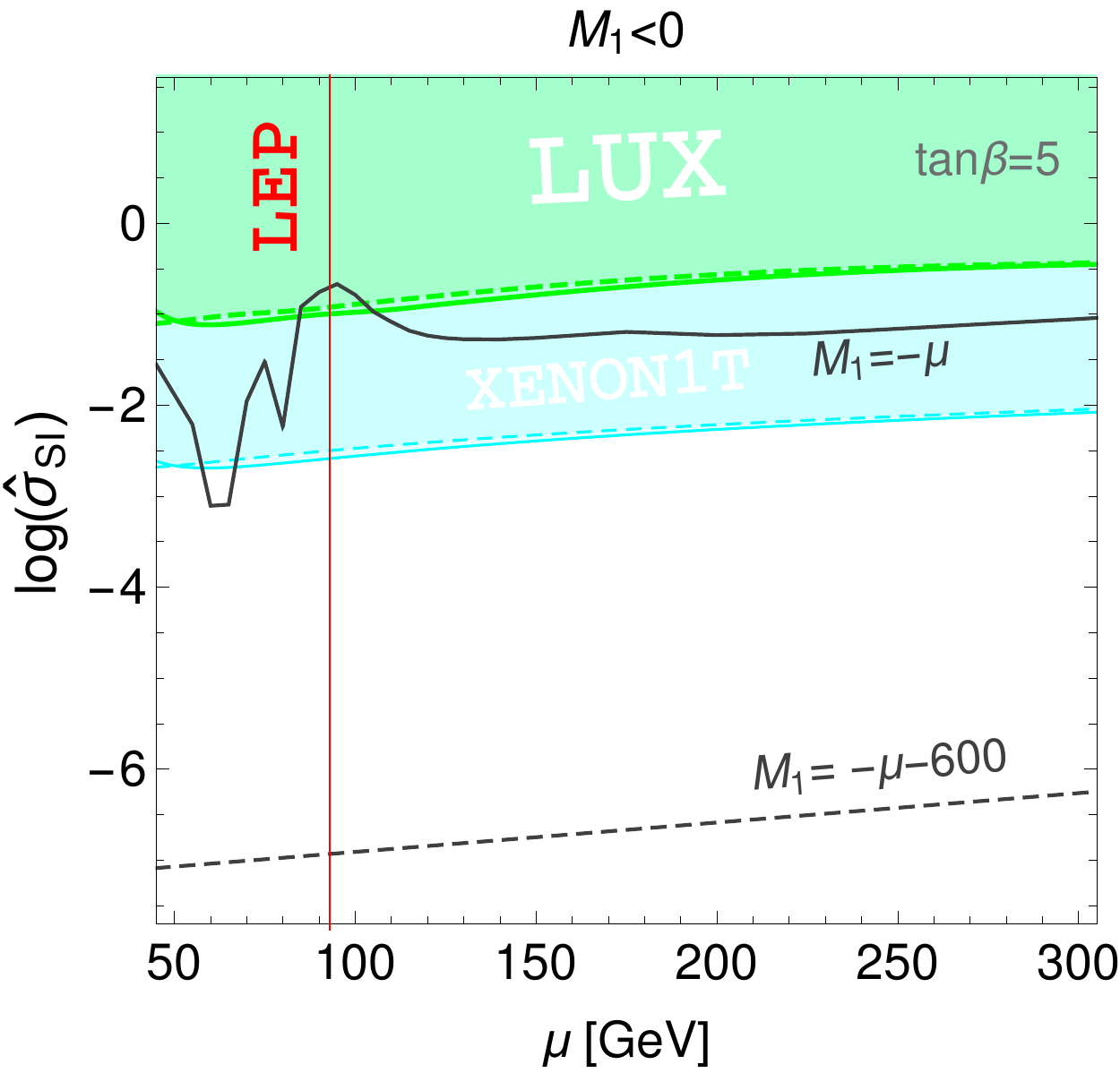}\\
\includegraphics[width=0.49\textwidth]{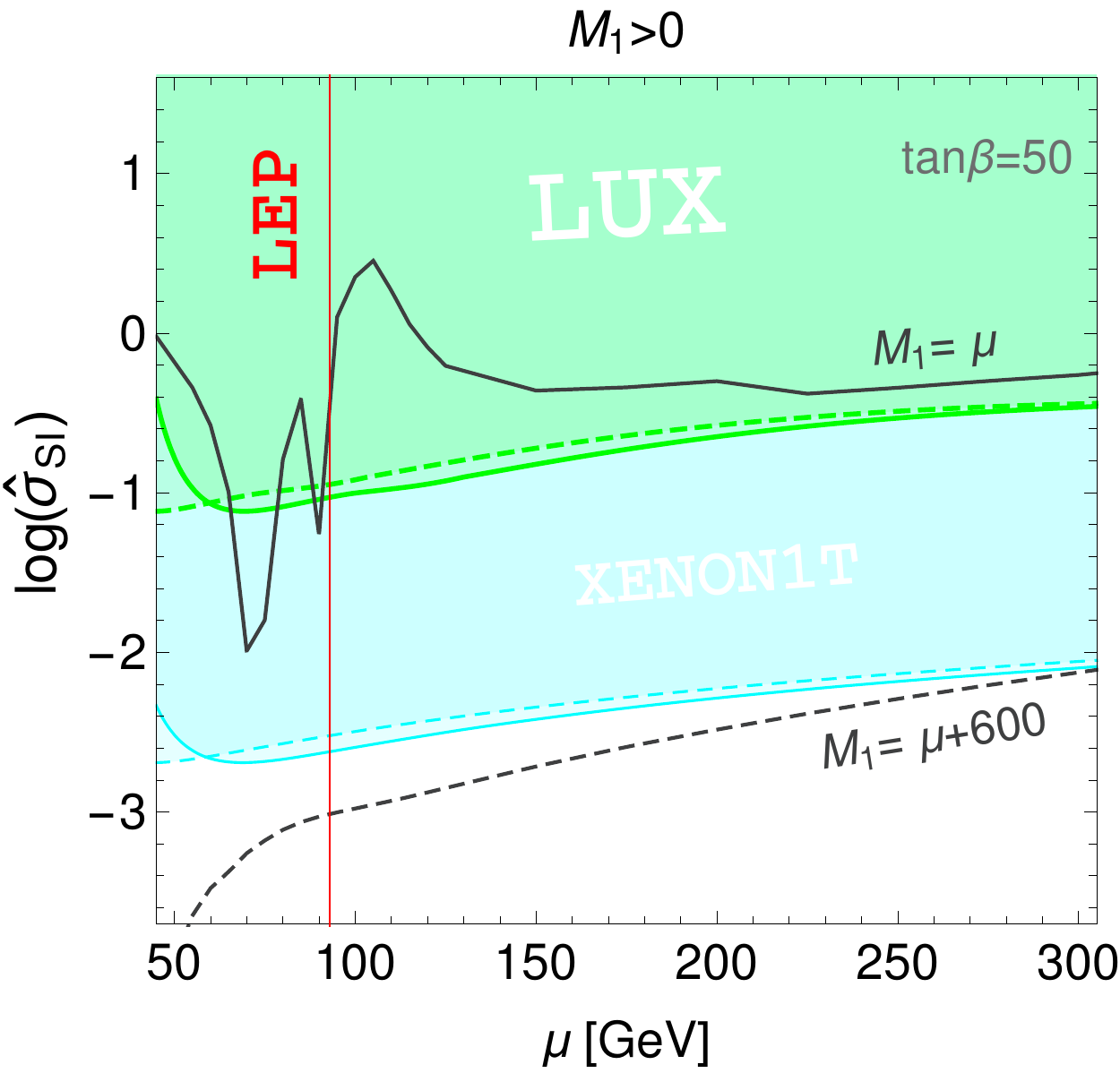}\hspace{.5cm}\includegraphics[width=0.49\textwidth]{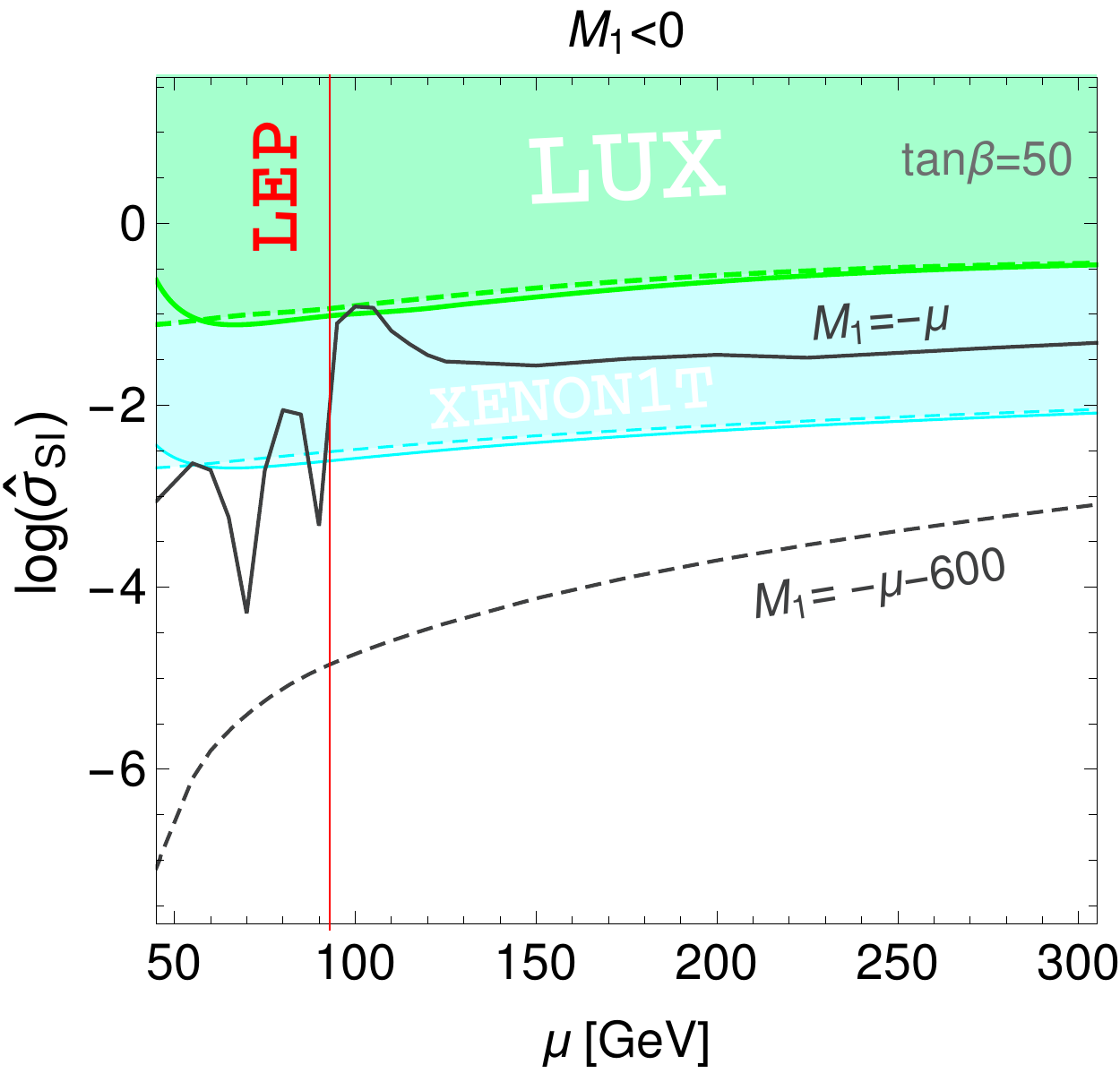}
\caption{The logarithm of the predicted value of the spin-independent annihilation cross section for DD $\hat{\sigma}_{\rm SI}=R_\Omega\,\sigma_{\rm SI}/(10^{-8}\,{\rm pb})$, 
rescaled by $R_\Omega/(10^{-8}\,{\rm pb})$ where $R_\Omega=\Omega_{\rm DM}/\Omega^{\rm Planck}_{\rm DM}$, is shown as a function of $\mu$ for $\tan\beta=5,50$ and positive (left) or negative (right) values of $M_1$ as indicated. The excluded limit from LUX (green), as well as the projected exclusion from XENON1T (cyan) are also shown for comparison, where the solid and dashed lines represent the exclusions for $|M_1|=\mu$ and $|M_1|=\mu$+600 GeV respectively.} 
\label{fig:sigmaSI}
\end{figure}

\begin{figure}[h!]
\includegraphics[width=0.49\textwidth]{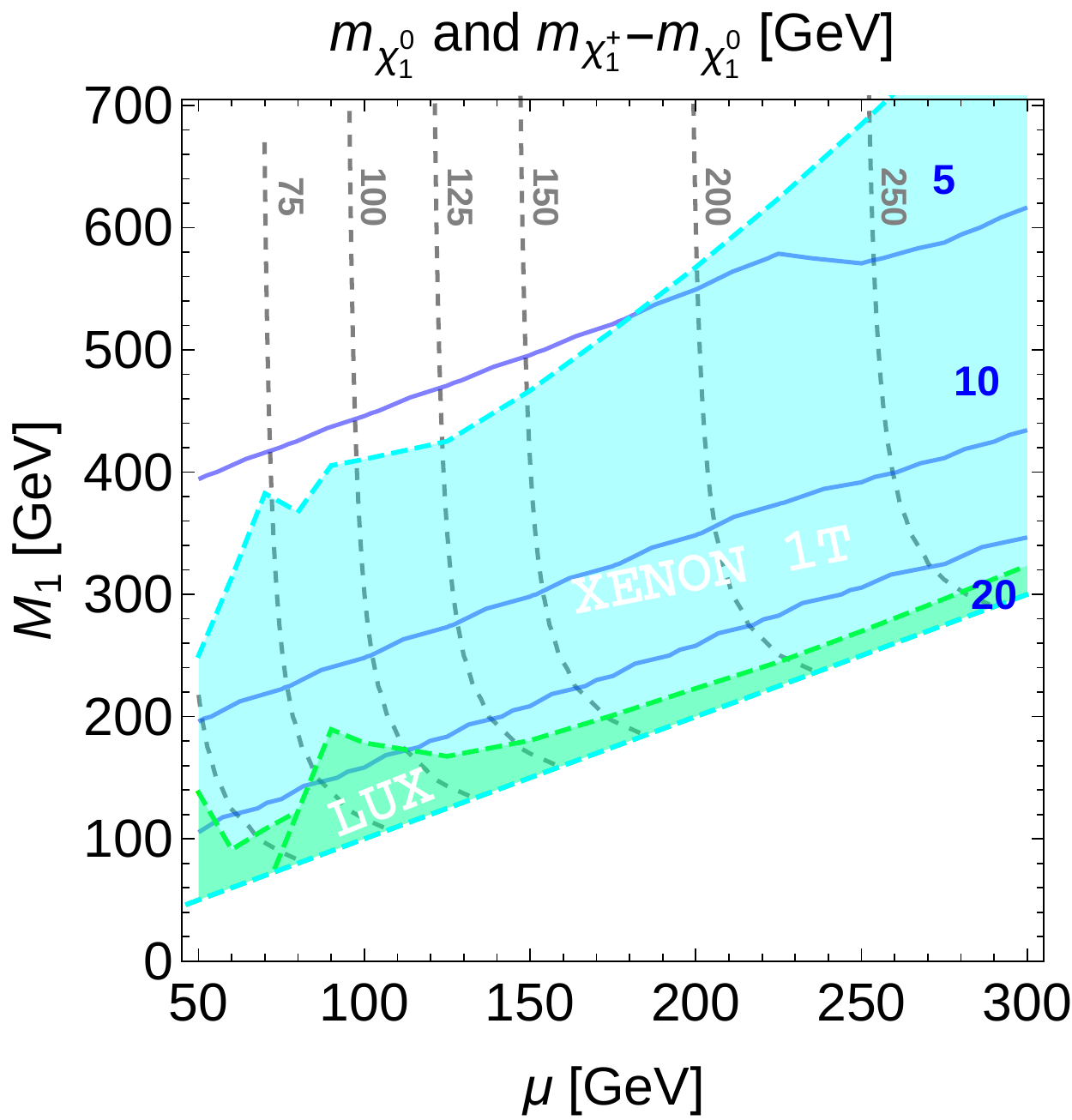}\hfill
\includegraphics[width=0.50\textwidth]{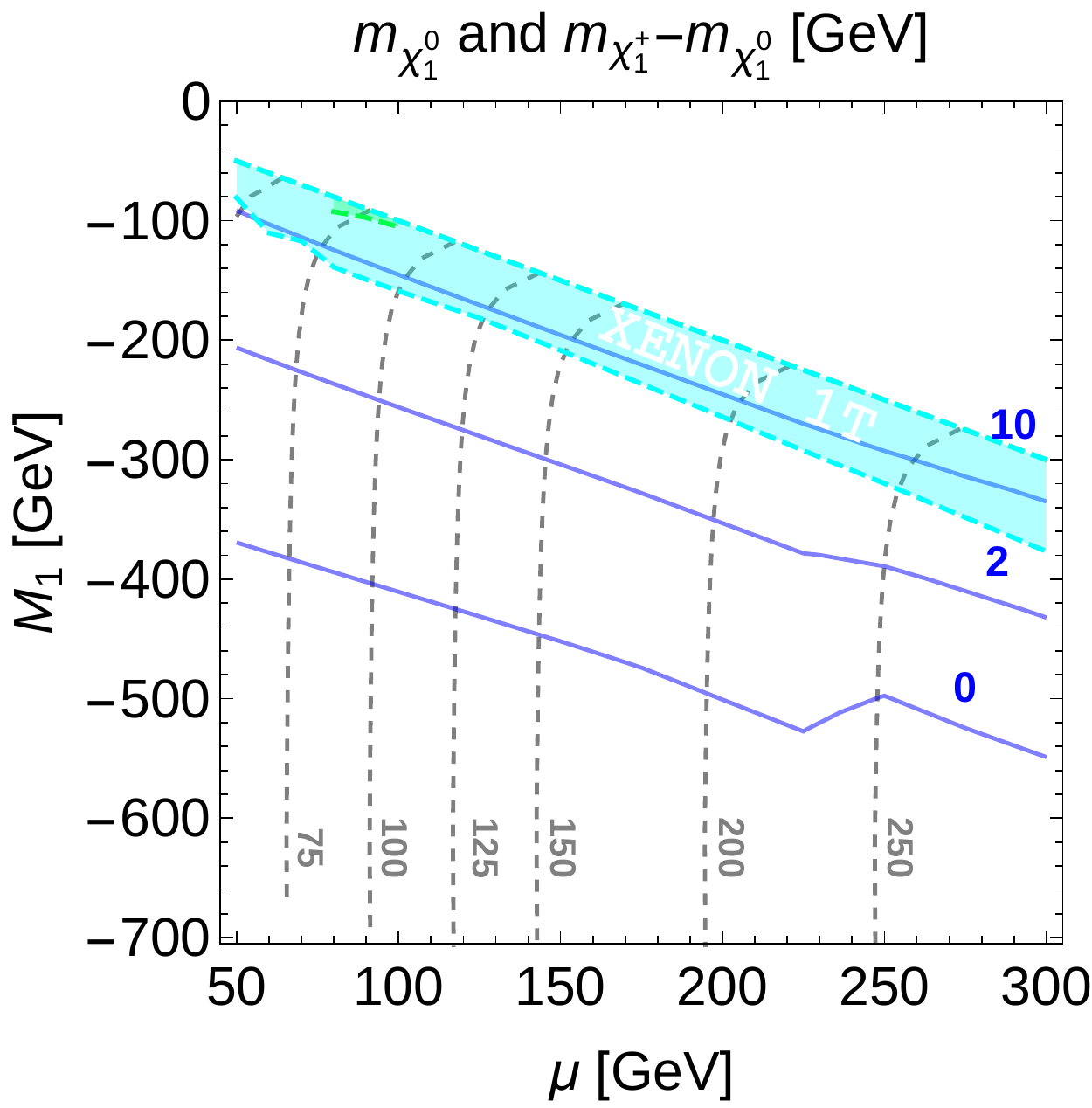}\hfill
\caption{The mass splitting between $\charpm{1}$/$\neut{1}$  (blue solid line) and 
mass of $\neut{1}$ (black dashed line) is shown in the $\mu$-$M_1$ plane, along with the region excluded by LUX results (green).}
\label{fig:mdm-mu-m1-2}
\end{figure}

In order to assess the compatibility of the scenarios under our investigation with existing experimental limits, we have evaluated  $\Omega_{\rm DM} h^2$, 
the spin-independent annihilation cross section ($\sigma_{\rm SI}$)
and the respective DD rates using {\texttt{micrOMEGAs 2.4.1}}~\cite{Belanger:2006is, Belanger:2010gh}.
In Fig.~\ref{fig:Omega} we show the results for $\Omega_{\rm DM} h^2$ as a function of
$\mu$ for  $\tan\beta=5,50$ and positive or negative values of $M_1$. Note that the uncertainty on $\Omega_{\rm DM} h^2$ is not shown, the full one-loop corrections are not yet available, but we expect that these are not too large and will not qualitatively change our conclusions.  
From these plots we see, as expected, that in general $\Omega_{\rm DM} h^2$ lies below $\Omega_{\rm DM}^{\rm Planck} h^2$, and decreases as $\neut{1}$ becomes increasingly higgsino-like.
This is because of the mass splitting between $\charpm{1}$ and $\neut{1}$, which becomes larger as $M_1$ decreases. This suppresses the coannihilation channels which otherwise lead to an efficient reduction of the relic density. Below the $WW$ threshold, the annihilation of the mixed gaugino-higgsino neutralinos therefore occurs via the Higgs and Z-bosons. The small bottom Yukawa coupling and the suppressed coupling to the $Z$ implies that this mechanism is not efficient, apart from at the $Z$ and $h$ resonance. 
Therefore the spike in the relic density $\mu\sim 100$ GeV can be explained by the fact that this is just below the $WW$ threshold. At lower values of $\mu\sim 70, 90$ GeV one moreover observes two dips corresponding to the Z-boson and Higgs funnels. This is most pronounced for positive values of $M_1$ and lower values of $\tan\beta$ where the mass splitting is larger.

In Fig.~\ref{fig:sigmaSI} we further show the spin-independent annihilation cross section for DD, 
again for positive and negative $M_1$ as in Fig.~\ref{fig:Omega}, where
instead of $\sigma_{\rm SI}$ we plot the rescaled quantity $R_\Omega\,\sigma_{\rm SI}$ (pb), where the scaling factor
$R_\Omega=\Omega_{\rm DM}/\Omega^{\rm Planck}_{\rm DM}$ allows easy comparison with the most recent 
limits (also reported in these plots) from LUX~\cite{Akerib:2013tjd}, as well as the projected limits from 
XENON1T after 2 years live-time and 1 ton fiducial mass (see e.g.~Ref.~\cite{Aprile:2012zx}), which in general assume the relic density to be the value measured by Planck.
Fig.~\ref{fig:sigmaSI} illustrates that the region with a low LSP masses and a higher mass splitting between $\charpm{1}$ and $\neut{1}$ (see Fig.~\ref{fig:mdm-mu-m1}), which is also easiest to see at
colliders, is in fact excluded as the $\sigma_{\rm SI}$ for the DD experiments is too high.
In the following we will further highlight the interesting complementarity between the reach of the collider searches and the DD searches, particularly interesting for low DM masses.
In Fig.~\ref{fig:mdm-mu-m1-2} we also show the mass splitting between $\charpm{1}$ and $\neut{1}$ in the $\mu$-$M_1$ plane of Fig.~\ref{fig:mdm-mu-m1}, along with the region excluded by LUX and the projected exlusion regions from XENON1T.
This emphasises that the region where $\mu$ and $M_1$ are very close each other is already excluded by LUX, which in turns puts an upper bound on the splitting between $\charpm{1}$ or $\neut{2}$ and $\neut{1}$. Note that for positive values of $M_1$, the splitting is larger, as is the mixing between bino and higgsino component of the DM. As larger mixing leads to larger couplings to Higgs bosons, DD is more sensitive to the case $M_1>0$.

A related question is whether these scenarios could be excluded by indirect detection (ID) experiments, i.e.~the detection of  
energetic $e^{\pm}$, $\gamma$, $p$ or $\bar{p}$, which may be created by the pair annihilation of weakly interacting massive particles (WIMPS).
It turns out that the strongest bounds on neutralinos coming from such experiments are set by gamma ray telescopes: both the Fermi-LAT gamma-ray space telescope~\cite{Ackermann:2011wa} as well as ground based telescopes. Fermi-LAT is sensitive to gamma rays particularly in the low mass range  up to $\mathcal{O}(100\,\mathrm{GeV})$. It is therefore particularly sensitive to lighter mixed bino-higgsino neutralinos, but the bounds are not competitive with those coming from DD.
 Here we are interested in light higgsinos with possibly some bino component, for which the relic density is in general much below the value measured at Planck/WMAP (see Fig.~\ref{fig:Omega}). In this case, all bounds from ID must be scaled by the square of the ratio of the predicted relic density to the experimental value. The rescaling appears as the pair annihilation cross section depends on the square of the local WIMP abundance. In \cite{Baer:2013vpa,Arrenberg:2013rzp} it was shown that the Fermi-LAT limits derived for WIMP annihilations into $WW$ (of which a large component of the total annihilation cross section should be comprised) are not yet sensitive to higgsino LSPs in the 100-350~GeV range mainly due to the predicted under-abundance of neutralinos in these scenarios.
 Note that  WIMPS having masses in the range 200 GeV to a few TeV will be probed by the future CTA array~\cite{Consortium:2010bc,Doro:2012xx}, i.e.~any higgsino or wino-like LSP for which the relic abundance is within an order of magnitude of Planck would be seen. In \cite{Cahill-Rowley:2014boa} a comprehensive scan of the pMSSM was carried out, and DM limits (both present and projected) coming from the LHC, DD and ID experiments were studied (note that the collider study was not dedicated to the region of interest of the present paper; our optimisation of the kinematical cuts goes beyond previous analyses). Ref.~\cite{Cahill-Rowley:2014boa} found an impressive degree of complementarity: in the range a few hundred GeV to 1 TeV DD poses stronger bounds, and for higher masses ID is more sensitive. Therefore we conclude that for the mass range studied in this paper, the LHC and DD limits are the most relevant.


%% file: 04_LHC.tex
\section{LHC potential to probe NSUSY}

In the previous section we have discussed the current and future sensitivity of underground experiments to 
NSUSY scenario. In this section we explore the LHC potential to probe NSUSY and demonstrate that it plays a crucial complementary role.

\begin{figure}[htbp]
\centering
\includegraphics[width=\textwidth]{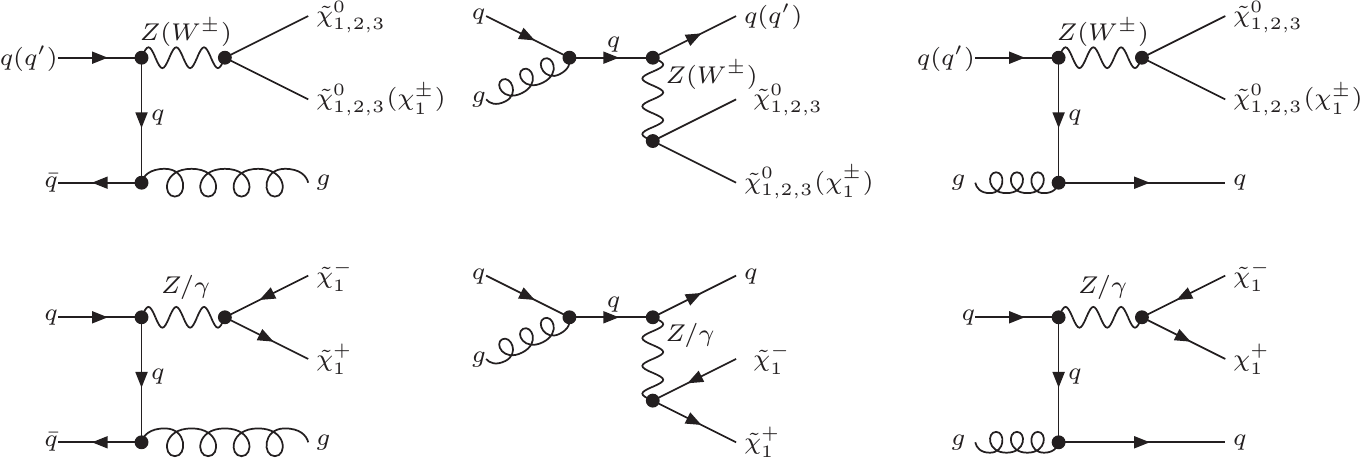}
\caption{\label{monojetFD} Representative diagrams for
pair neutralino-chargino production in association with quark/gluon 
leading to monojet signature.}
\end{figure}

The LHC's most sensitive searches for quasi-degenerate  $\charpm{i}$ and $\neut{i}$ scenario
are monojet signatures, i.e.~the production of a pair of electroweakinos
 through the s-channel exchange of a SM EW gauge boson,  $Z$, $\gamma$ or $W^\pm$, accompanied by hard QCD initial state radiation (ISR) via the process
\begin{equation}
p p  \to \chi_a \chi_b j \quad \chi_{a,b}=\neut{1,2,3},\charpm{1} \ .
\label{eq:ino_pp}
\end{equation}
The corresponding Feynman diagrams can be found in  Fig.~\ref{monojetFD}.

As we discuss below, the main problem for the signal search in this channel is the large background, the dominant contribution coming from the EW production processes $Z$+jets and $W$+jets.
We begin our analysis of the monojet signal from the recasting of the LHC  at $\sqrt{s}=$~8~TeV data and the respective experimental results, and then analyse the prospects of the 13~TeV LHC run with both standard and high luminosity (HL) options.
Both ATLAS~\cite{ATLAS:2012zim} and CMS~\cite{CMS:rwa} have performed studies of monojet signatures at the LHC Run1, which have been interpreted  in the context of an EFT approach.
However this approach cannot be used for the NSUSY scenario because of the $Z, W,\gamma$ mediating interactions as indicated  in Fig.~\ref{monojetFD}.
Hence, in the case of NSUSY,
\bea
\textrm{DM with EW mediators } \slashed{\Rightarrow} \,  {\cal L}_{eff} = \frac{1}{\Lambda^2} (q \Gamma \bar q)  \, ( \chi \bar \chi ) 
\eea
where $\Gamma$ is some Lorentz structure and $\chi$ is the DM particle. 
Therefore it is necessary to recast the searches in terms of this NSUSY scenario and not to use limits from $\Lambda$-${\rm DM}_{\rm mass}$ plane.


\subsection{Analysis Setup}

In this section we describe the different aspects of our simulation of both the signal and the most important backgrounds, implementing the important steps of hadronization and fast detector simulation.

We performed a parton-level simulation using MadGraph v1.5.11 ~\cite{Alwall:2011uj} 
with the MSSM model available on the FeynRules web page \cite{Christensen:2008py} implemented in UFO format \cite{Degrande:2011ua}\footnote{\tt{http://feynrules.irmp.ucl.ac.be/wiki/MSSM}}, and cross-checked results against  CalcHEP~\cite{Belyaev:2012qa} with the MSSM model from the HEPMDB website\footnote{\tt{http://hepmdb.soton.ac.uk/hepmdb:0611.0028}}.
 
At the level of matrix-element  we have generated the  production of a pair of electroweakinos via the s-channel exchange of a SM EW gauge bosons,  accompanied by hard QCD initial state radiation.
Parton level SM background simulations have been also cross checked between two packages. 
Our choice of PDF sets is CTEQ6L1~\cite{Pumplin:2002vw} and we used the MadGraph dynamical choice of renormalization scale which is equal to the geometric mean of $\rm{Mass^2} + P_T^2$ for the final state particles.
 Parton showering, hadronisation and decay of the unstable particles were simulated using 
PYTHIA v6.4 \cite{Sjostrand:2006za} while detector effects have been simulated with Delphes3 \cite{deFavereau:2013fsa} employing a suitable CMS card.
Finally, the background processes yields, which include $Z+j$, $W+j$, $t\bar t$,
QCD and single top processes, have been taken from the experimental results for 8 TeV analysis and has been simulated using
MadGraph+PTYHIA+DELPHES chain for the 13 TeV analysis.
\begin{figure}[htbp]
\begin{center}
\includegraphics[width=0.7\textwidth]{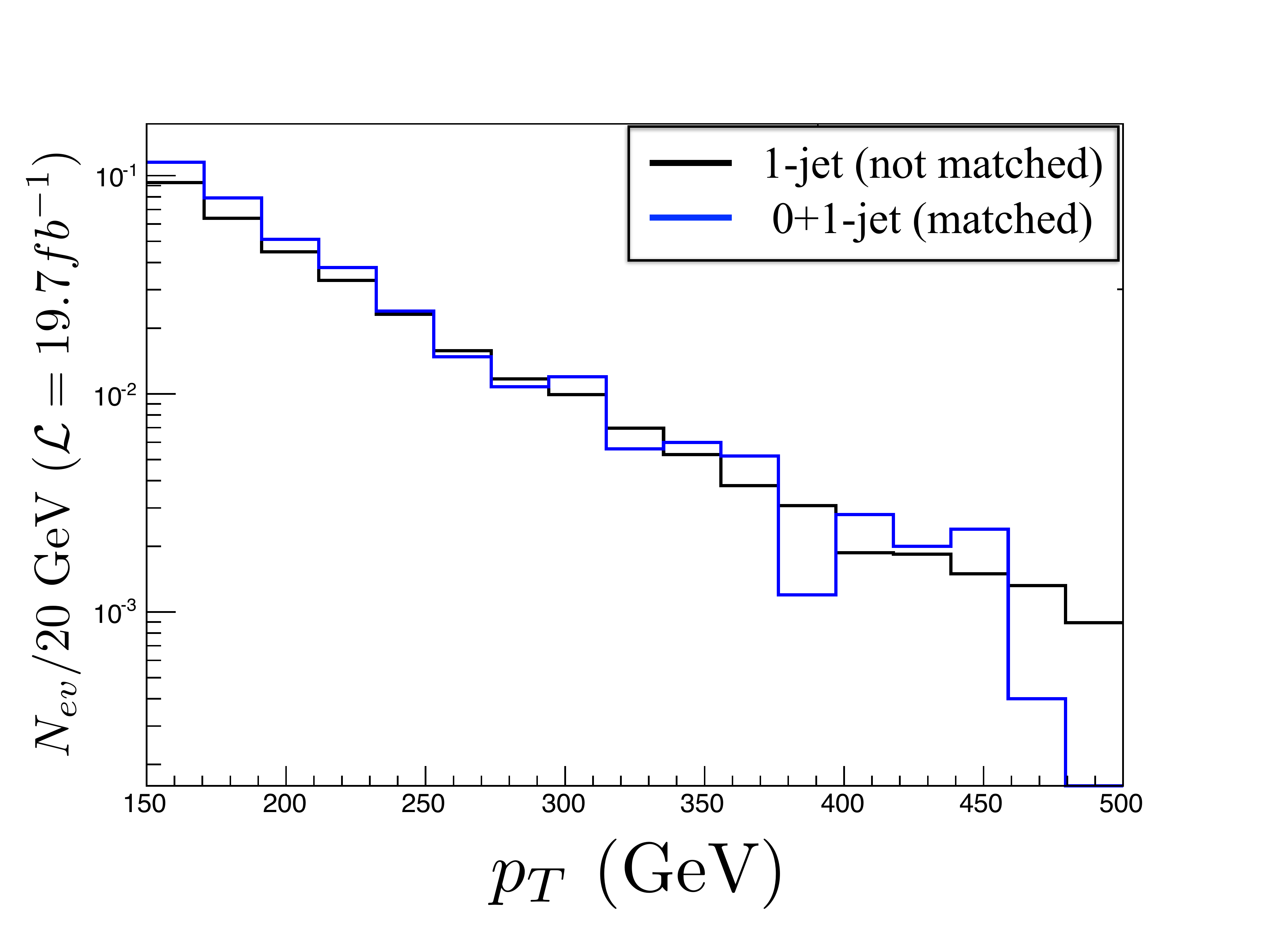}\hfill
\caption{Leading jet $p_T$ cross section distributions for the case of the 0+1 jet matched sample (blue) and 1 jet sample (black) for the $\neut{1}\neut{2}$ production at the 8 TeV LHC. }
\label{fig:mathing_plot}
\end{center}
\end{figure}

For the signal and background events at the parton level
we have applied a cut on the jet transverse momentum of $p_T^j>90$~GeV 
at the generation level. This is a subtle point.
One might argue that, since dealing with processes involving QCD radiation, it is necessary to
apply a merging procedure between the hard jet generated via the parton-level matrix element and the soft jets generated by the showering algorithm, merging therefore the 0-jet and 1-jet samples.
However, since  a hard jet is selected at the analysis level (e.g.~$p_T^{j}> 110$ GeV for the 8 TeV LHC analysis and higher for the 13 TeV case, with a final selection requirement of a high $\met$, somewhat correlated with the jet $p_T$~\cite{Brooijmans:2014eja}) 
we found that this matching was unnecessary, and generated  just the one jet sample.
Moreover, avoiding matching in this case allowed us to make our analysis much more effective and obtain enough statistics in the  high $p_T^j$ region.
To illustrate the validity of this procedure,  in Fig.~\ref{fig:mathing_plot} we present the $p_T$ distribution of the leading jet for the case of $\neut{1}\neut{2}$ production.
Shown in blue is the leading jet $p_T$ for the case of a 0+1 matched jet sample, while the case of the 1-jet events being unmatched is shown in black.
One can indeed see that in case of a high $p_T$ cut on the leading jet at the analysis level,
the matched and unmatched distributions of the leading jet
are very similar, indicating that  contribution from the  0-jet matched sample is negligible,
and that the $p_T$ of the leading jet is dominated by the one-jet sample in the high $p_T$ region.

\subsection{LHC Run1: the reach of monojet searches}

We start our analysis with the exploration  of the 8 TeV LHC potential to probe NSUSY and the recasting of the 
respective experimental results.
For this purpose we have chosen  CMS monojet analysis~\cite{CMS:rwa} which has been done for  the data recorded at $\sqrt{s}=8$ TeV with  19.5~fb$^{-1}$
integrated luminosity\footnote{The respective ATLAS analysis, which leads to similar results, can be found in ~\cite{Aad:2015zva}.}.
We have applied  the following trigger selection followed by the   cut-flow according to the CMS analysis:
\begin{itemize}
\item
Two triggers which require $\met>120$~GeV or
$\met>105$~GeV and a jet with $p_T>$~80 GeV and within $|\eta|<$ 2.6.
\item
The analysis then requires that the jet with the highest transverse momenta has $p_T>$~110 GeV and
$|\eta|<$ 2.4.
\item
Events with more than two jet with $p_T>$ 30 GeV and $|\eta|<$ 4.5 are discarded together with events
where $\Delta\phi(j_1,j_2)<$~2.5, where $j_1$ and $j_2$ are the leading and sub-leading jets, to reduce QCD background.
\item The $W$ production background was suppressed by applying a veto on events with one electron or muon
satisfying $p_T>$ 10 GeV cut and events with one tau jet with $p_T>$~20 GeV and $|\eta|<$ 2.3.
\item
Finally the analysis was performed in 7 regions with an increasing requirement of $\met$: $\met>$ 250, 300, 350, 
400, 450, 500 and 550 GeV.
\end{itemize}

We have then derived the signal significance for each signal region from the number of signal events (S),
after having imposed the cuts above, and the number of background events (B), using the following expression  as in the CMS analysis:
\bea
\alpha = 2(\sqrt{S+B}-\sqrt{B}) \ ,
\eea
which is similar to the more common $S/\sqrt{S+B}$ for $S\ll B$, but is more robust with respect to downward fluctuations~\cite{Bityukov:1998zn}.

\begin{figure}[h!]
\includegraphics[width=0.5\textwidth]{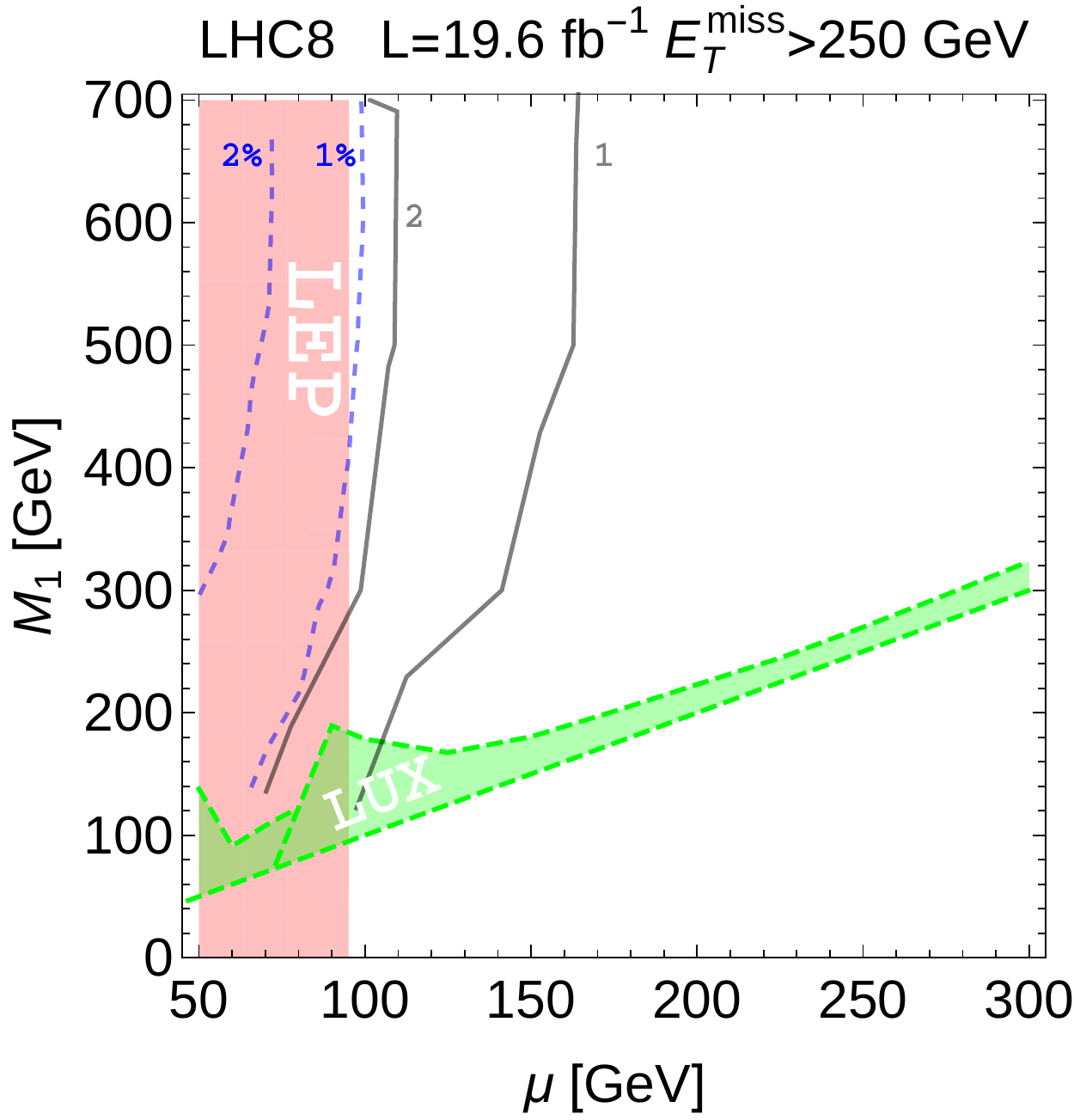}\hfill
\includegraphics[width=0.5\textwidth]{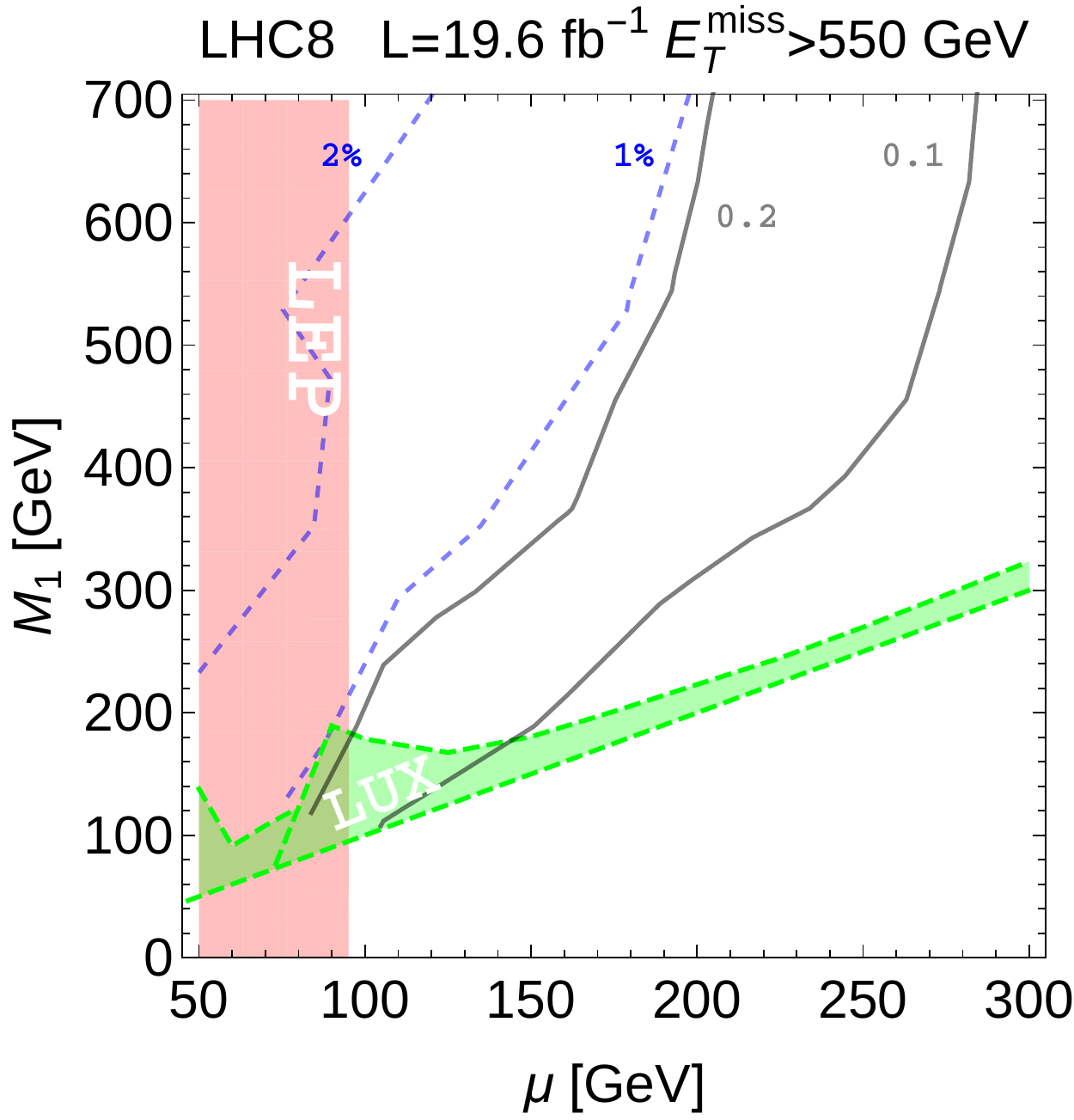}
\caption{Contours for the $2(\sqrt{S+B}-\sqrt{B})$ (gray) and S/B (blue-dashed)in the plane ($\mu$,$M_1$) for the signal regions 
1 and 7 as defined in~\cite{CMS:rwa}. LUX and LEP exclusions are shown in green and red.
 }
\label{fig:SoversqrtSpB_Etmiss_i}
\end{figure}
In Fig.~\ref{fig:SoversqrtSpB_Etmiss_i} we show contours of iso-significance $\alpha$ (solid gray) in the $\mu$-$M_1$ plane for the lowest (left frame) and highest (right frame) requirements on $\met$. The LUX exclusion is further shown by the green shaded area as well as the LEP2 limit on charginos by the red shaded area. In addition, the S/B ratio is shown by a blue-dashed line.
By {\it only} inspecting significance contours, one {\it could think} that the Run1 LHC data could slightly extend the LEP2 limit using the low $\met$ signal region:
the area on the left of the lines of $\alpha$=2 would be ruled out at 95\% confidence level (CL). 
However, one then observes that the $S/B$ never goes above 2\% in the region allowed by LEP2, 
while the actual systematic uncertainties of this analysis are of the order of 5-10\% ~\cite{ATLAS:2012zim,CMS:rwa}. This means that the low  $S/B$ ratio for the 8 TeV LHC is the main obstacle to going beyond the LEP2 limits.
We can also  see from Fig.~\ref{fig:SoversqrtSpB_Etmiss_i} (right) that the higher $\met$ cut increases
the S/B ratio, eventually at the expense of the signal, meaning that the significance
drops below $\alpha$=2 level in the parameter space allowed by LEP2.
Therefore we conclude that {\it Run1 LHC does not have the potential to test the NSUSY scenario beyond the LEP2 limit due to systematic uncertainties.}

\subsection{13 TeV LHC  potential and complementarity to underground experiments}

As we have seen in the previous section,
the LHC Run1 is not sensitive to the NSUSY parameter space we consider 
due to the low statistical significance and the fact that the low S/B ratio remains below the systematic errors. 

In this section we study the LHC Run2 case and show that the higher collider luminosity and energy
allow us to choose kinematical cuts, bringing the S/B ratio to a desirable level
while keeping the statistical significance at a high enough level in order to establish sensitivity to  the NSUSY parameter space.
Being a very challenging scenario, we make projections up to HL configuration of the LHC machine for different assumptions regarding the $\met$ cuts and the control over systematic uncertainties.

\begin{figure}[h!]
\epsfig{width=0.5\textwidth,file=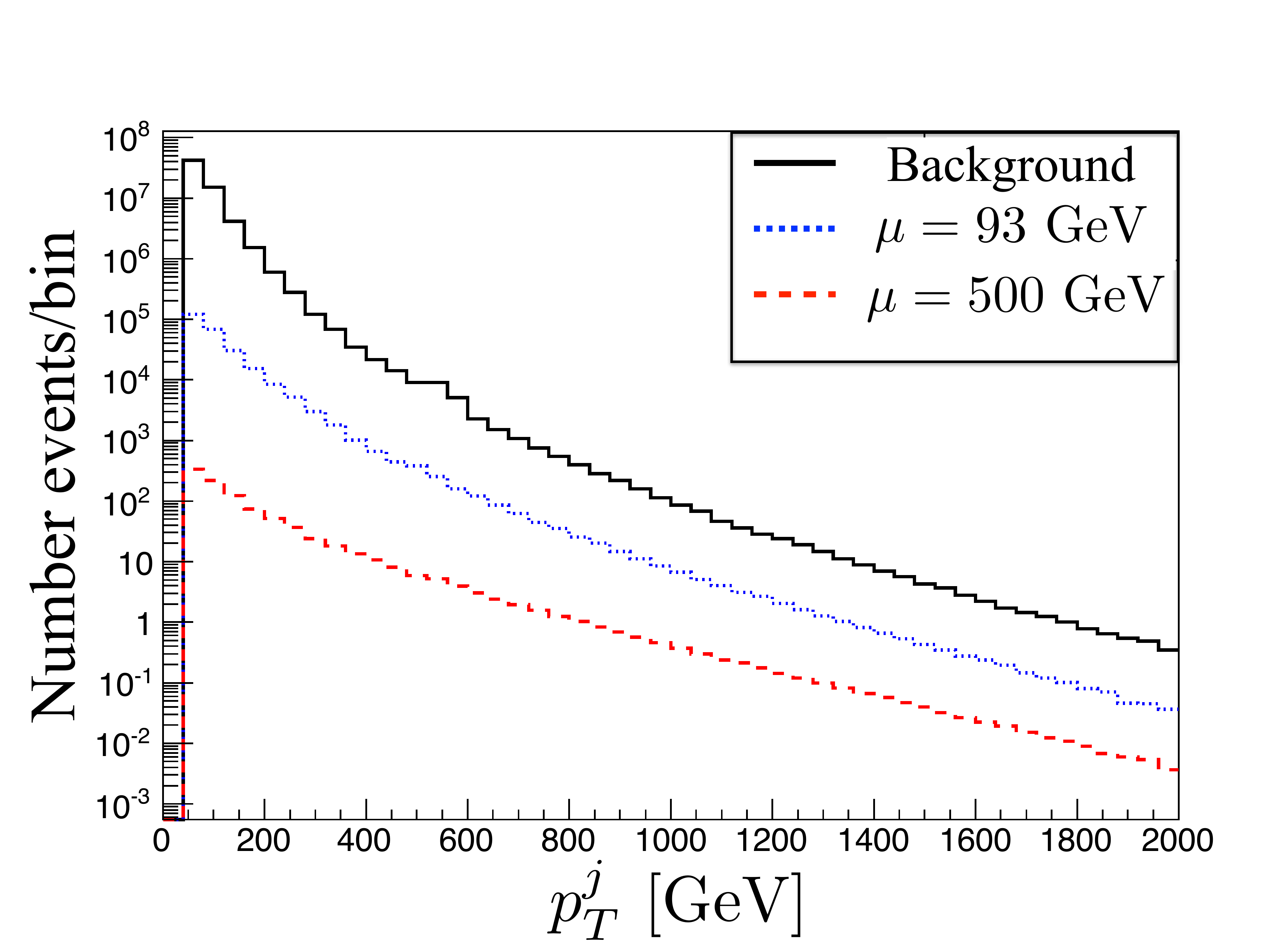}%
\epsfig{width=0.5\textwidth,file=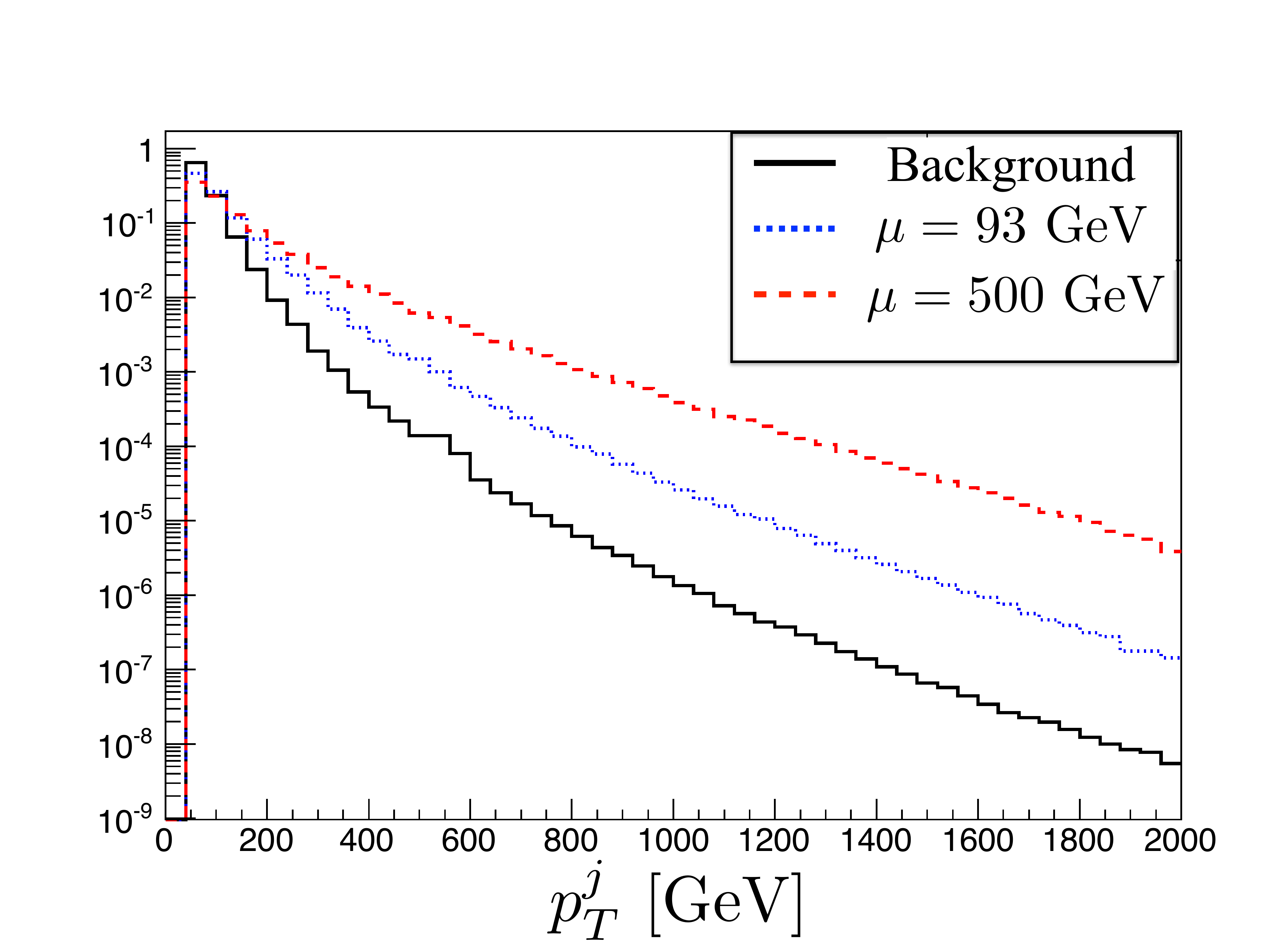}
\caption{Signal (dotted blue and dashed red) and $Zj$ background (solid black) parton-level
$p_T^j$ distributions for the 13 TeV LHC for the NSUSY
scenario. Left: $p_T^j$ distributions for  100 fb$^{-1}$ integrated luminosity.
Right: normalised signal and  $Zj$ background distributions.
}
\label{fig:signal-vs-bg}
\end{figure}

In order to ensure that the S/B ratio is under control, we compare the relative size and shape 
difference of the signal versus the dominant irreducible background $Z+jet \to \nu\bar{\nu}+jet$ ($Zj$).  The relevant parton level distribution are shown as a function of the jet $p_T$ in Fig.~\ref{fig:signal-vs-bg}.
One can see that even for $\mu=93$~GeV corresponding to $m_{\chi^0_1}\simeq$~100~GeV,
the background is about 3 orders of magnitude higher than the signal for a low $p_T^j>$ cut.
An important feature of the signal versus background is that the $shape$ of the background distribution is quite different from the signal: the background falls more rapidly with $p_T^j$, and the difference in the slope with respect to the signal is bigger for higher neutralino masses. The different in slope is  mainly due to the mass difference  between the neutralino, from the signal, and the neutrino, from the background.
 One should also notice that the difference between the shapes of the signal and background $p_T^j$ distributions vanishes for very large values of $p_T^j \gg m_{\chi^0_1}$, as one would expect.
This provides us with our main strategy to optimise the LHC sensitivity to the NSUSY parameter space, that is to find the maximal value for $\met$ cut while maintaining the statistical significance at high enough level.
On inspecting this figure one can already see that in order to achieve S/B ratio at about 5\% level the  value $\met$ cut should be around of 1~TeV.

The signal process analysed is the same as that in Eq.~(\ref{eq:ino_pp}), with the only difference being that the initial requirement on the QCD ISR has been increased to 150 GeV, motivated by our preliminary study in Ref.~\cite{Brooijmans:2014eja}.

Along with the signal processes, we have simulated the two main backgrounds
\begin{equation}
\begin{split}
& p p \to Z j\to \nu_l\bar\nu_l j \\
& p p \to W j\to l^+\nu_l+c.c.
\end{split}
\end{equation}
with $l=e,\mu,\tau$.

Inspired by the 8 TeV CMS monojet search we have then applied the following cut-flow
\begin{itemize}
\item We require a leading jet with $p_T>$ 200 GeV and $|\eta|<2.4$
\item We apply a veto on events with more than two jets with $p_T>$ 30 GeV and $|\eta|<$4.5
\item We require $\Delta\phi(j_1,j_2)<$2.5
\item We apply a veto on electrons and muons with $p_T>$ 10 GeV.
\item We apply a veto on taus with $p_T>$ 20 GeV and $|\eta|<2.3$.
\end{itemize}
We have then defined signal regions with increasing cuts on $\met$; an example cut flow is provided in Table~\ref{cutflow}.
We have not generated $t\bar t$, QCD and single top background, though we have applied to our simulated samples cuts that reduce 
these background to a negligible level with respect to $Wj$ and $Zj$, see details in Refs.~\cite{ATLAS:2012zim,CMS:rwa}.

\begin{table}
\begin{center}
\begin{tabular}{|c|c|c|| c| c| }
\hline
					 & $Zj$, $Z\to\nu\nu$	& $Wj$, $W\to l\nu$ 	& $\mu=$ 100 GeV        & $\mu=$ 200 GeV \\
					 &                      &              		& $M_1$=700 GeV	        & $M_1$=800 GeV \\
\hline
\hline
 Initial  \# of events   	   		 	 & 3.15$\cdot 10^6$	&  1.25$\cdot 10^7$ 	& 3.63$\cdot 10^5$ 	& 6.45$\cdot 10^3$ \\
\hline
 $p_T^j>$ 200 GeV $|\eta^j|<2.4$           & 1.05$\cdot 10^6$	&  4.11$\cdot 10^6$	& 1.73$\cdot 10^5$	& 3528	\\
 Jet veto 				 & 8.7$\cdot  10^5$	&  3.13$\cdot 10^6$	& 1.33$\cdot 10^5$	& 2691	\\
 $\Delta\phi(j_1,j_2)<2.5$		 & 7.2$\cdot  10^5$	&  2.3 $\cdot 10^6$	& 1.10$\cdot 10^5$	& 2320	\\
 Veto $e^\pm,\mu^\pm,\tau^\pm$ 		 & 7.2$\cdot  10^5$	&  6.8 $\cdot 10^5$ 	& 1.08$\cdot 10^5$	& 2301 \\
 $\met>200$ GeV  		 & 6.4$\cdot  10^5$	&  4.3$\cdot  10^5$ 	& 9846 			& 2188 \\
 $\met>600$ GeV  		 & 4353			&  1002		 	& 171 			& 93 \\ 
 $\met>700$ GeV  		 & 1703			&  250		 	& 80 			& 47 \\  
 $\met>800$ GeV 		 & 694			&  0		 	& 37			& 22 \\
\hline
\end{tabular}
\caption{Cutflow for the two main SM background and two choices of signal for the 13 TeV LHC with 100 fb$^{-1}$ of integrated luminosity.
The initial number of events corresponds to $p_T^j>150$~GeV cut.
}
\label{cutflow}
\end{center}
\end{table}

\begin{figure}[h!]
\includegraphics[width=0.52\textwidth]{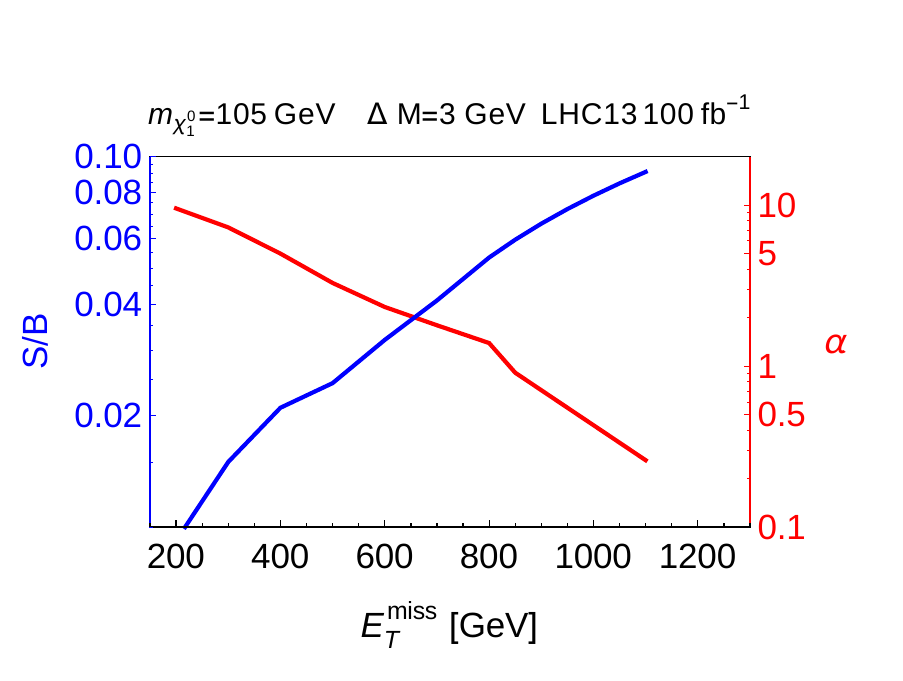}%
\includegraphics[width=0.52\textwidth]{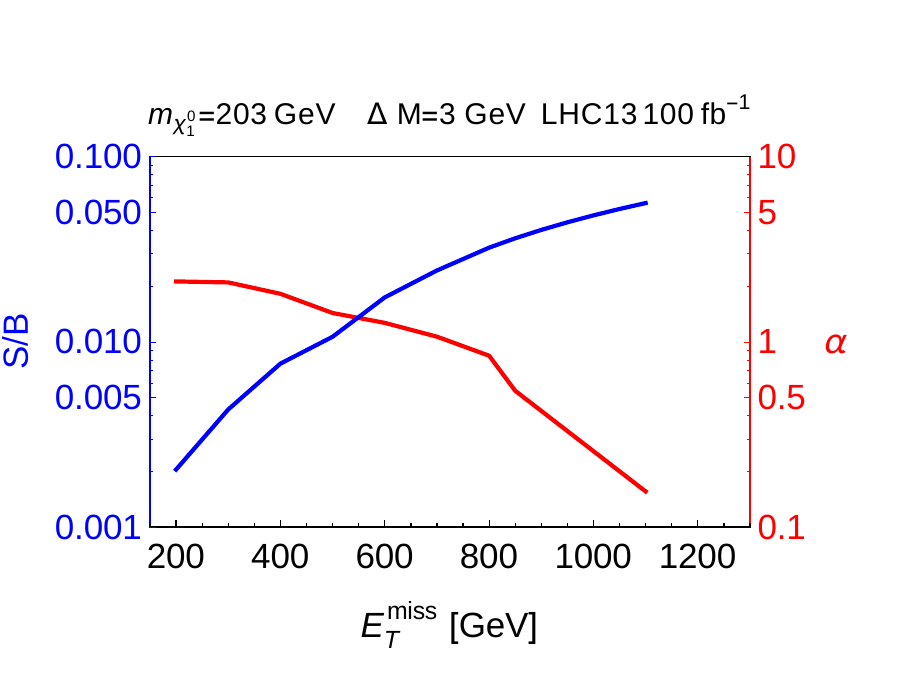}
\caption{S/B and $\alpha$ as a function on the final selection cut on the $\met$  for $\neut{1}\sim$ 100 GeV (left panel) and $\sim$ 200 GeV (right panel).}
\label{fig:SoB_sign_MET}
\end{figure}

As observed for the case of 8 TeV LHC, a strong tension arises in attempting to simultaneous maximise 
the S/B ratio  and the statistical significance.
We demonstrate this in Fig.~\ref{fig:SoB_sign_MET}, where we plot S/B and $\alpha$ as a function on the final selection cut on the $\met$ for both the case of  $\neut{1}\sim$ 100 GeV (left panel) and $\sim$ 200 GeV (right panel).
This figure clearly indicates  that in order to achieve a high enough S/B ratio in keeping with the expected level of systematic uncertainties, a hard cut on $\met$ ought to be applied, 
which at the same time pulls the significance down, due to the reduction in the number of signal events.

Given this tension it is therefore important to optimize the $\met$ cut to provide a high enough S/B ratio and to keep $\alpha>$2 (5) in order to obtain an exclusion (discovery).
The NSUSY parameter space can be conveniently described  and presented in $m_{\neut{1}}$-$\Delta$M
plane, where for $\Delta$M we have chosen the mass difference between the lightest chargino and the DM particle. In this plane, for a given value of integrated luminosity, the optimal $\met$ cut   can be chosen by the point where S/B and $\alpha$ cross or are as close to each other as possible.
This is related to the fact that the iso-significance contours are shifted to the {\it left} in the $m_{\neut{1}}$-$\Delta$M plane with the increase in the $\met$ cut due to the {\it decrease} of signal statistics, while iso-S/B contours are shifted to the {\it right} at the same time due to the {\it increase} of S/B ratio. 
Therefore the case when the respective iso-contours cross/are close to each other, would  provide the {\it maximal}  exclusion or discovery area in  the $m_{\neut{1}}$-$\Delta M$ plane.

\begin{figure}[htbp]
\centering
\includegraphics[width=0.6\textwidth]{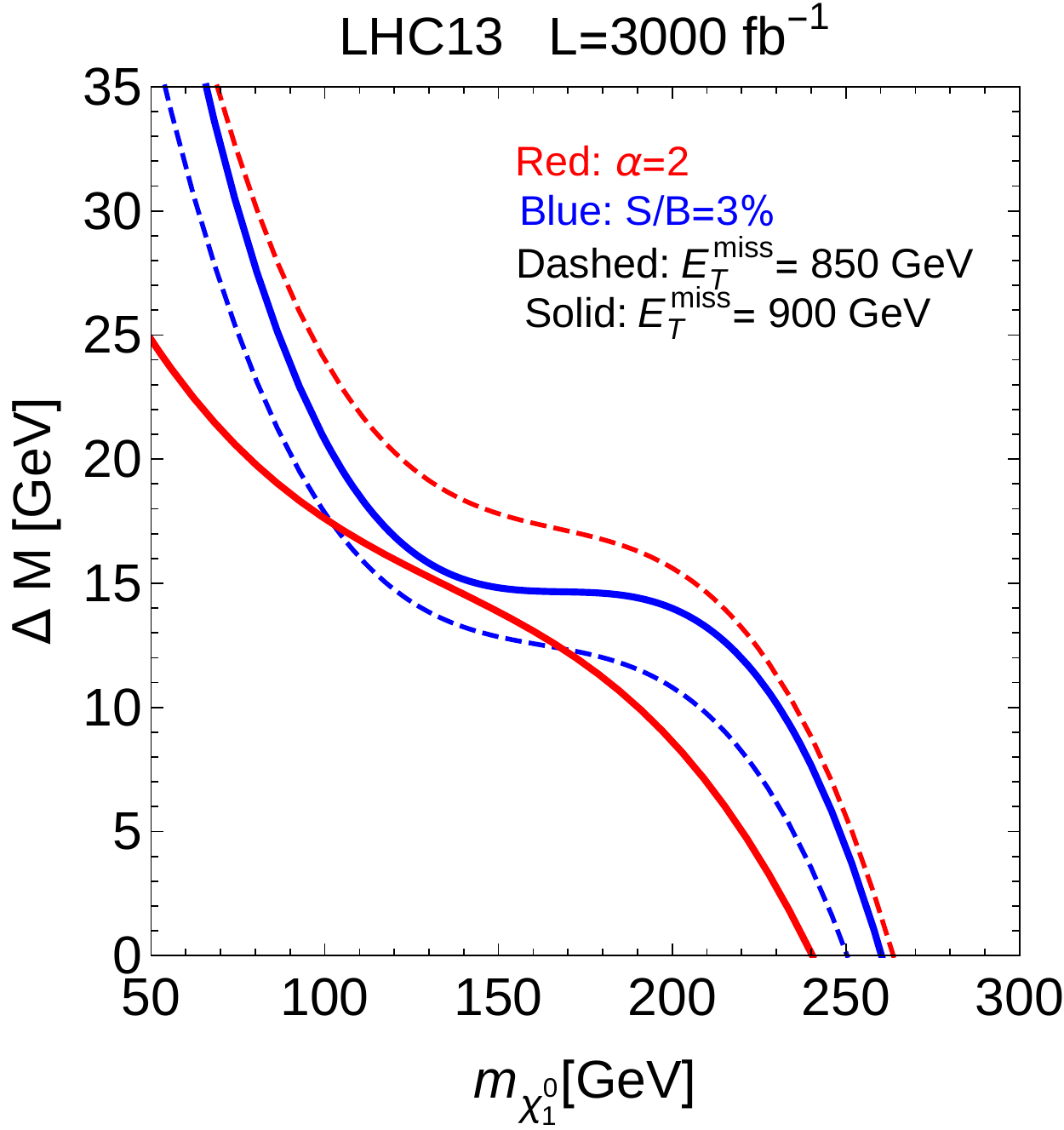}%
\caption{S/B (blue) and $\alpha$ (red) isocontours for two choices of $\met$ cut: 850 GeV (dashed) and 900 GeV (solid) in the $m_{\chi^0_1}$-$\Delta $M plane.}
\label{sb-signif-tension}
\end{figure}

We illustrate this in Fig.~\ref{sb-signif-tension}
which presents S/B and significance isocontours in the $m_{\neut{1}}$-$\Delta$M
plane for two different cuts on $\met$, 850 and 900 GeV.
One can see that indeed for $\met>850$~GeV, the exclusion area
is below  $S/B=3\%$ (blue dashed) contour, 
while for  $\met>900$~GeV, the area below  $\alpha=2$ (red solid) contour
is excluded. Since for the first case the exclusion area is bigger,
the $\met>850$~GeV is better choice for the optimal cut.
 We have found that a cut around 600 (850) GeV for 100 fb$^{-1}$ (3000 fb$^{-1})$  provides
 $\alpha\simeq 2$ and $S/B\simeq 0.03$ iso-contours optimally close to each other, which maximises the reach of the 13 TeV LHC for the NSUSY parameter space.
 The proximity of $\alpha\simeq 2$ and $S/B\simeq 0.05$ iso-contours eventually requires higher $\met$ cut which is found to be around 950 GeV, 
 and as a result leads to a poorer 13 TeV LHC reach as we discuss below.
While presenting results in the $m_{\neut{1}}$-$\Delta M$
plane, we separate the cases  $M_1 > 0$ and $M_1 < 0$,  
which differ due to different bino-higgsino components and mass splittings as discussed earlier,
which eventually  becomes less and less relevant as we approach the low $\Delta$M region. 

\begin{figure}[htb]
\includegraphics[width=0.5\textwidth]{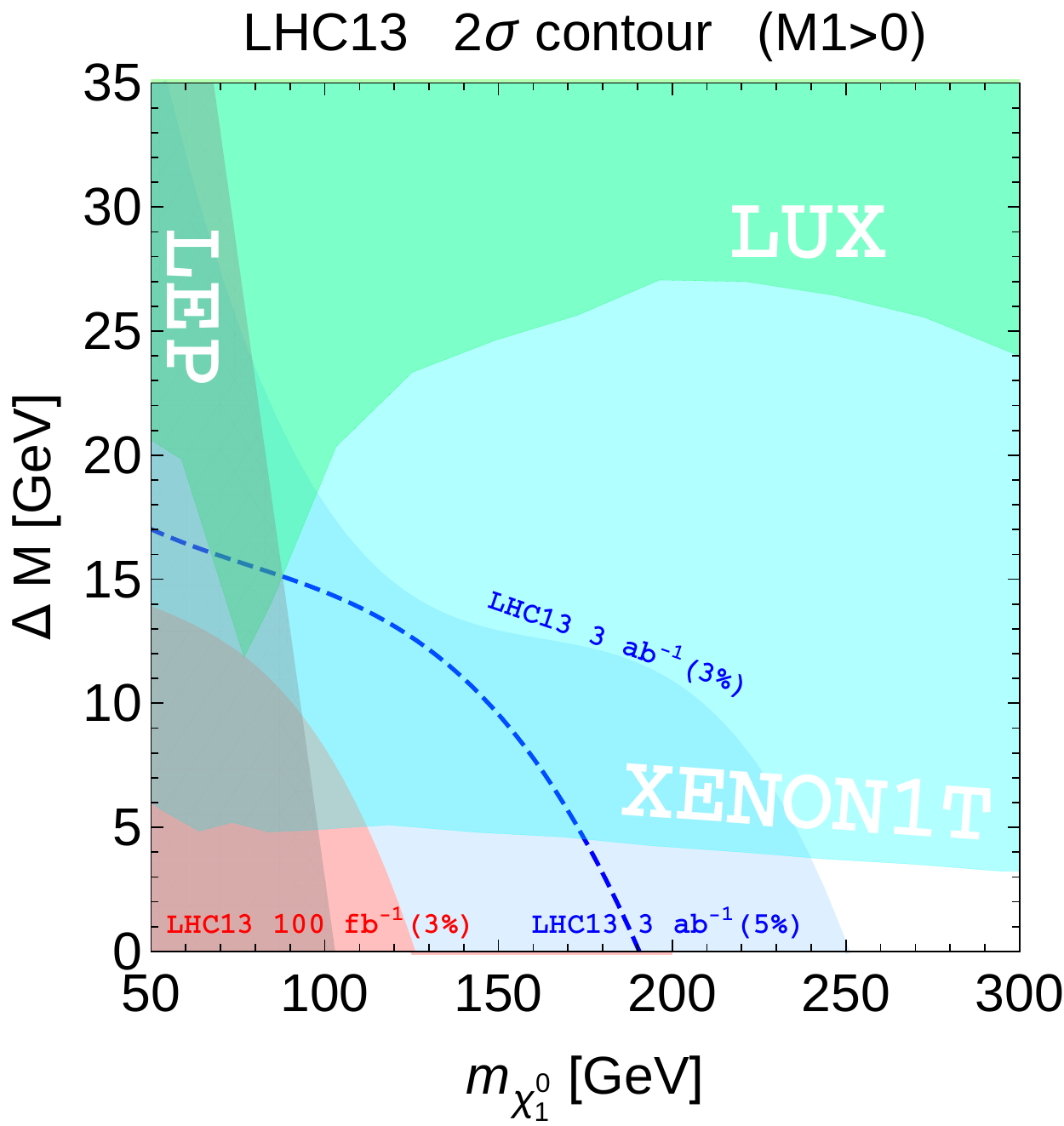}%
\includegraphics[width=0.5\textwidth]{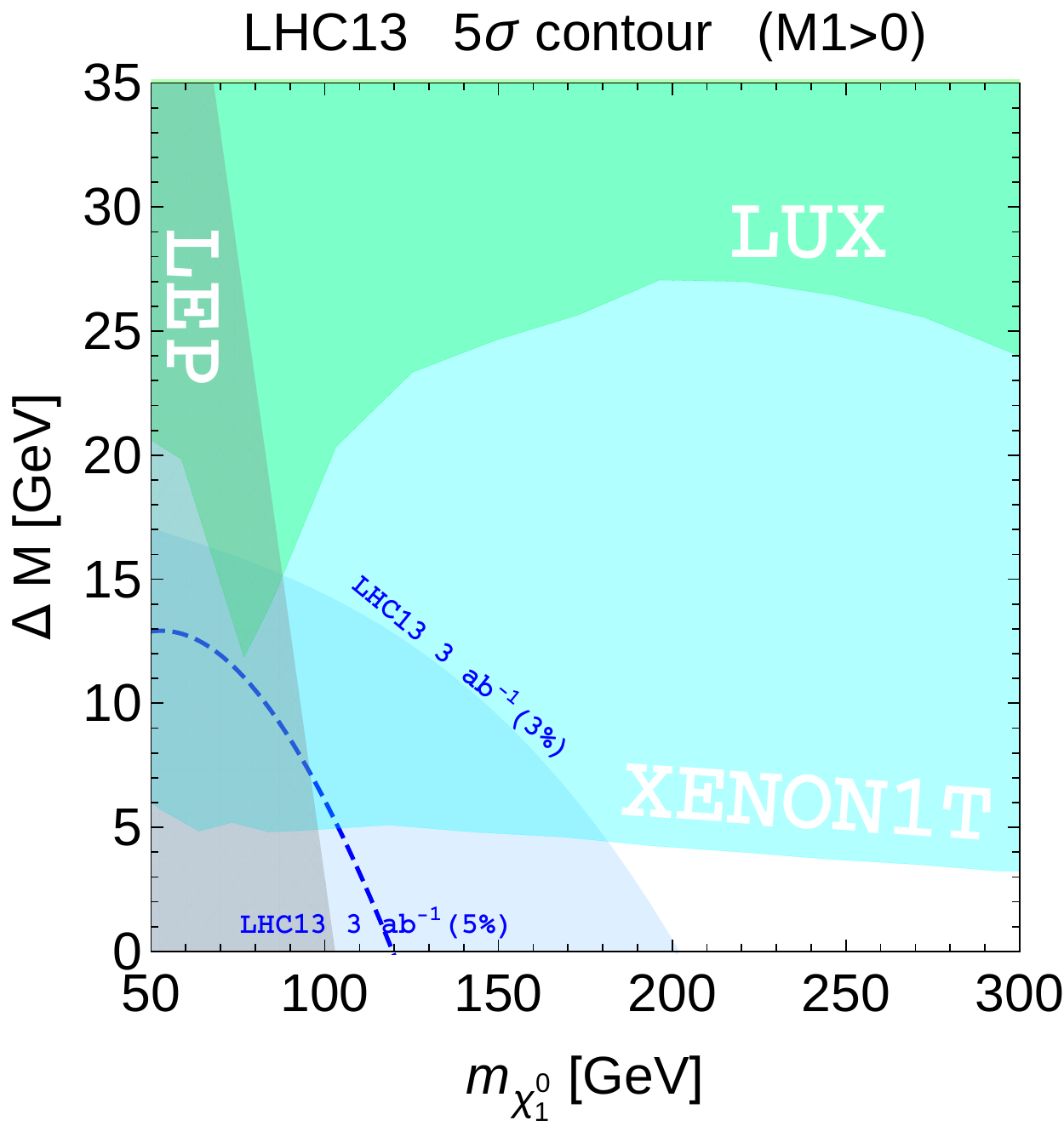}
\caption{Exclusion (left) and discovery (right) contour lines for the 13 TeV LHC at the end of the LHC Run2 (light red region) and of the HL-LHC (light blue region) assuming S/B$>$3\%. For the latter case also the case S/B$>$5\% is shown. The region excluded by LUX and the projected exclusion 
by XENON1T are also shown, together with the LEP limit on the $\charpm{1}$ mass. $M_1>\mu$ is considered here.}
\label{proj_m1p}
\end{figure}

\begin{figure}[htb]
\includegraphics[width=0.5\textwidth]{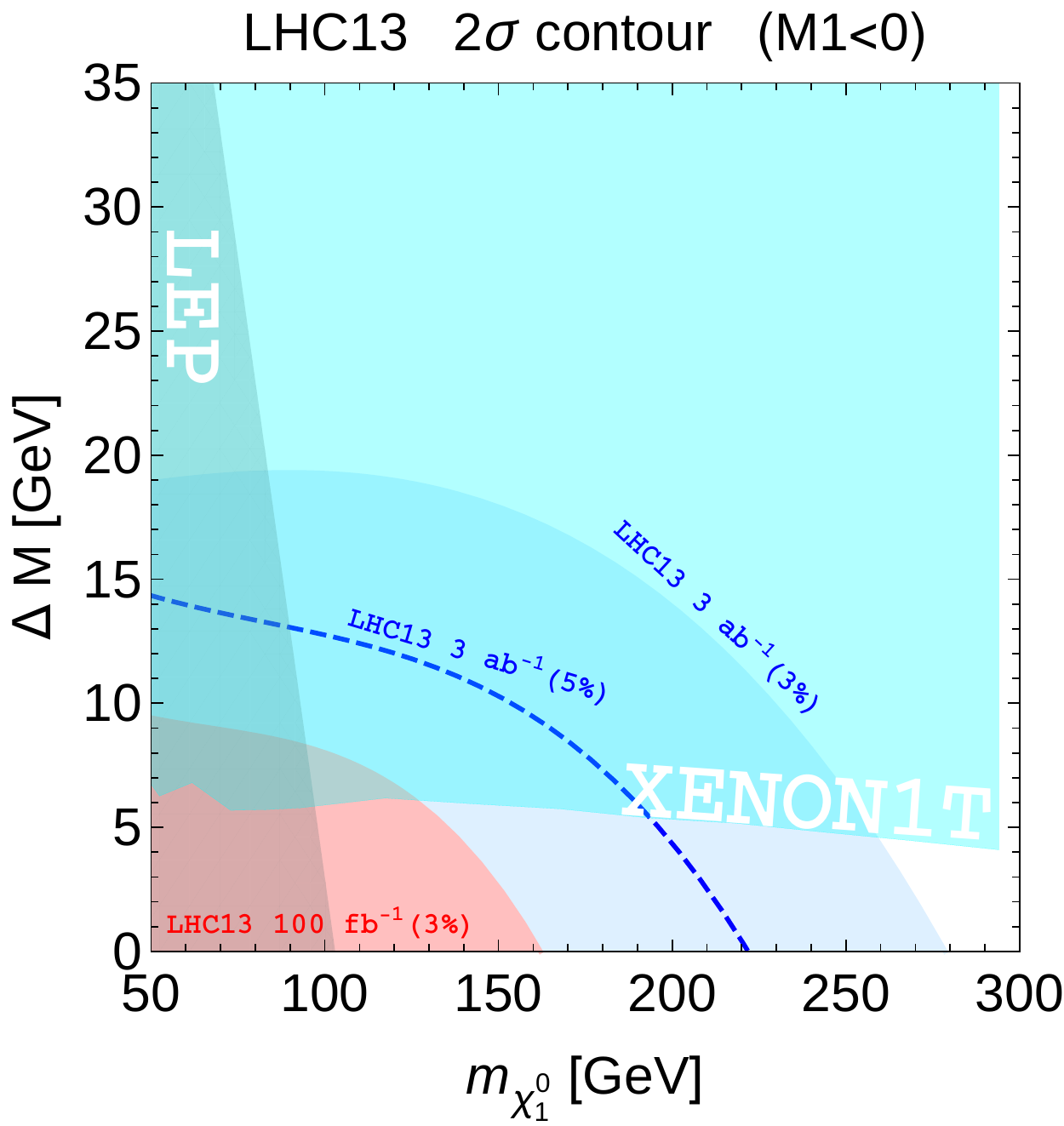}%
\includegraphics[width=0.5\textwidth]{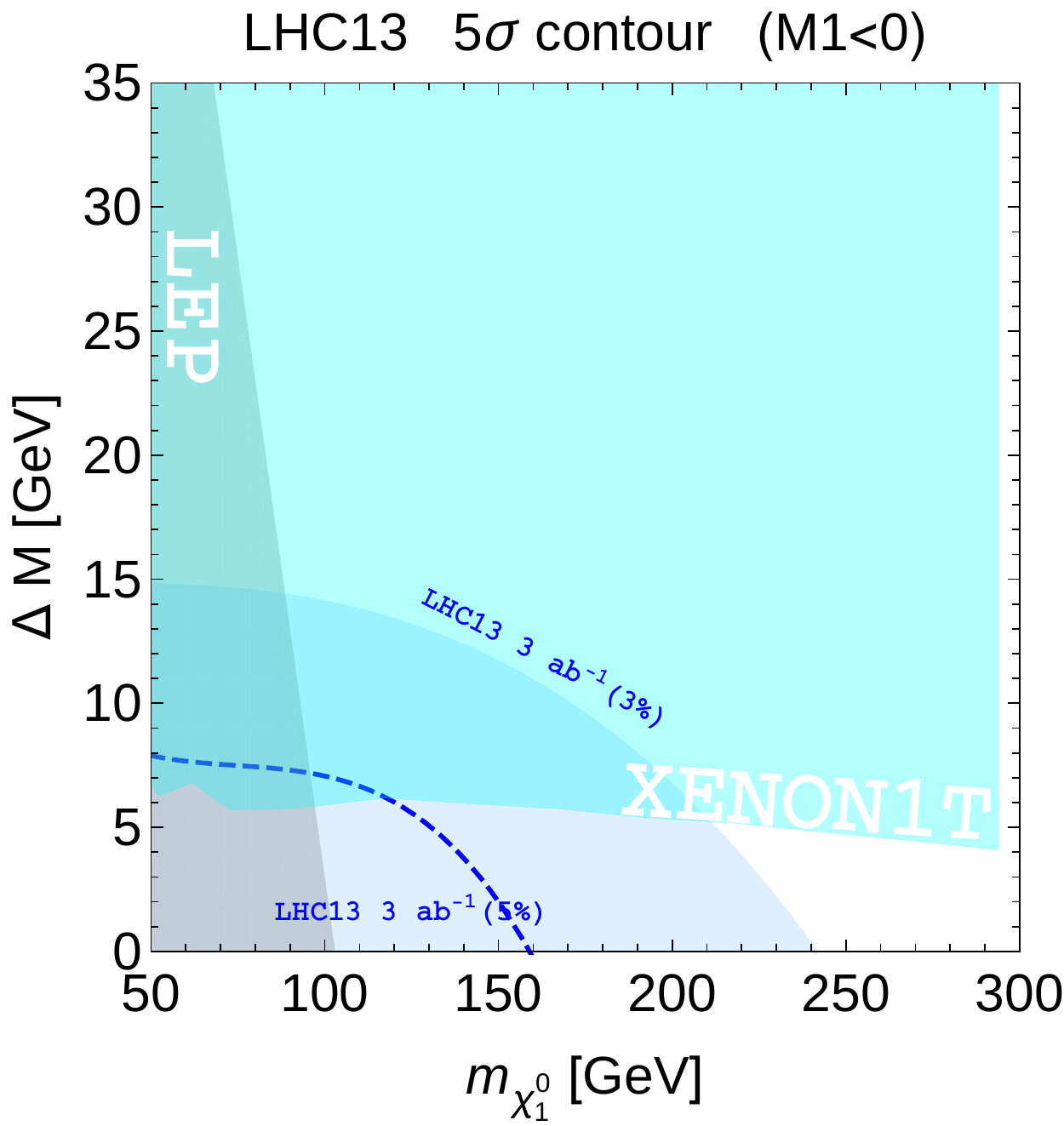}
\caption{Exclusion (left) and discovery (right) contour lines for the 13 TeV LHC at the end of the LHC Run2 (light red region) and of the HL-LHC (light blue region) assuming S/B$>$3\%. For the latter case also the case S/B$>$5\% is shown. The region excluded by LUX and the projected exclusion 
by XENON1T are also shown, together with the LEP limit on the $\charpm{1}$ mass. $M_1<-\mu$ is considered here.}
\label{proj_m1n}
\end{figure}

In Fig.~\ref{proj_m1p} and Fig.~\ref{proj_m1n} we show results for $M_1>\mu$ and $M_1<-\mu$ (note that we have chosen $\mu >0$). The left panels contain the 2$\sigma$ exclusion LHC reach while  the right panels contain the 5$\sigma$ discovery LHC potential.
Both the cases of requiring  3\% and 5\% for the S/B ratio are shown, the latter just for the High Luminosity (HL)-LHC scenario.
In  the same plot we present also LUX sensitivity and the projected sensitivity of  XENON1T to the
in the $m_{\neut{1}}$-$\Delta M$ plane.
One can see that at Run2, with $\sim$ 100 fb$^{-1}$ of integrated luminosity collected, the LHC will be able to exclude up to $\sim$150 GeV $\neut{1}$ with $\Delta M$ below 5 GeV, if the systematic error can be kept  at the  3\% level.
The LHC will therefore surpass the LEP sensitivity for this scenario which reached the limit $\charpm{1}>$ 103 GeV for low $\Delta$M  values as shown in Fig.~\ref{proj_m1p} and   Fig.~\ref{proj_m1n}.
By the end of the HL-LHC run, up to $\sim$ 250 (200) GeV LSP could be excluded for low mass splitting 
with a S/B$>$~3\%~(5\%).
It is quite remarkable that the LHC has the maximal sensitivity in the low $\Delta $M region. This reach
is nicely complemented by the LUX results and the projected exclusions for XENON1T, which 
 cover the region with higher mass splitting.
Such complementarity would allow LHC and DD experiments to completely exclude NSUSY
  scenario with DM mass up to $\sim$ 250 GeV.
Discovery prospects for this scenario are shown in the right panel of Fig.~\ref{proj_m1p} 
which demonstrates  that $\sim$ 180~(110) GeV $\chi^0_1$ can be discovered for S/B $>$~5\%~(3\%) at the end of the HL-LHC run, while with 100 fb$^{-1}$ the LHC will not have discovery sensitivity.

At the end of this section 
we would like to discuss a potentially promising new signature
involving a monojet plus soft di-leptons which was studied in terms of the
NSUSY parameter space in Ref.~\cite{Baer:2014kya}. In short, in this paper  the signal from the second neutralino decaying leptonically into the lightest neutralino has been studied. 
It was suggested that one could trigger on soft leptons 
and use the upper cut on di-lepton invariant mass
below 10 GeV to suppress background and extract the signal from compressed
$\neut{1}-\neut{2}$ production. After the suggested cuts, the background 
can be reduced down to about the 6 fb level bringing it below the signal,
which would allow one to claim a discovery for chosen benchmarks with only 
100 fb$^{-1}$ of integrated luminosity.
One should note however, that in spite of the comprehensive set of backgrounds,
$jb\bar{b}$ was not considered in this paper.
After application of the cuts from ~\cite{Baer:2014kya}
we have found that $jb\bar{b}$ background is in fact dominant for the 
monojet plus soft di-lepton signal and is about two orders of magnitude 
above the backgrounds taken into account in the aforementioned study.
An  estimation of  $jb\bar{b}$ background is not trivial,  
since the cross section of the $jb\bar{b}$ process is very high, about 100 nb
for soft  initial cuts on  jet $p_T$, while the efficiency  of the 
selection cuts is very low, about $10^{-5}$.
Therefore,  a reasonable estimate of this background requires either the simulation of at least  $10^6$
events or the extraction of this background from future experimental data.
Our preliminary results indicate that one should develop a dedicated strategy to suppress this $jb\bar{b}$ background in order to make the monojet plus soft di-lepton signal a viable tool 
for the exploration of the NSUSY parameter space.

%% file: 05_Conclusions.tex
\section{Conclusions}

In this paper we have explored the complementary  potential of the Large Hadron Collider and  underground experiments to probe  Dark Matter (DM) in the Natural Supersymmetry (NSUSY) scenario.
This study, which combines searches from different kinds of experiments, has to be done in the context of a specific model, as (model-independent) Effective Theory (EFT) approaches are very limited in scope, see e.g. the discussion in Refs.~\cite{Buchmueller:2013dya,Busoni:2013lha}.  
In particular the EFT approach is not applicable for well motivated NSUSY scenario, which we study here,
where DM has direct couplings to Standard Model electroweak (EW) gauge bosons and the Higgs.

Current limits on simple SUSY scenarios are at the TeV range, in clear tension with naturalness arguments and hence with 
the motivation for introducing SUSY in the first place. A possible explanation for this situation is that the manifestation of SUSY is not as simple as one expects, but there is more complexity in the structure of SUSY at high-energies. Notwithstanding, one would still expect that the particles more directly related to the tuning of the EW scale remain light in the spectrum. This leads to a generic expectation that DM in NSUSY should have a sizeable Higgsino component.
 
 While being theoretically attractive this scenario also represents a clear example of how colliders and underground experiments can complement each other. Indeed, while underground experiments have a larger mass sensitivity than colliders, being able to probe the multi-TeV region, colliders can cover parameter space hidden from DM direct detection (DD) experiments.
 Specifically, the increase of  the DM higgsino component makes NSUSY parameter space increasingly difficult to probe in DD experiments.
In this  region higgsino-like DM is quasi-degenerate with two other particles, the second neutralino and the lightest chargino,
and the increase of higgsino component is correlated with the  decrease of this mass splitting,  $\Delta$M.
 At the same time, the LHC  sensitivity increases
 with the increase of the higgsino component of DM and reaches its maximum for very low values of $\Delta$M.
We have conveniently described  the NSUSY parameter space in the $\mu$-$M_1$ region and have translated 
it into sensitivity of the LHC and DM DD experiments in the  $m_{\neut{1}}$-$\Delta$M plane,
where for $\Delta$M we have chosen the mass difference between the lightest chargino and DM.
 We have studied the current and the future reach of underground experiments in this  region, as well as given predictions for the relic abundance in NSUSY. 
 
We would like to point out that  we have presented combined LHC and DM DD  experiments results for the whole
$m_{\neut{1}}$-$\Delta$M NSUSY space, rather than for chosen benchmarks.
Moreover, we have optimised the final $\met$ selection cut to keep the S/B ratio at a high enough level, to deal with the systematic errors on the background, which aer one of the main problem for the exploration of the NSUSY parameter space.
As a result, we have found that the 8 TeV LHC unfortunately does not allow to test the NSUSY parameter space beyond the LEP2 limits, while the 13 TeV LHC at 3 ab$^{-1}$ has the potential to significantly surpass the LEP2 reach and to cover DM masses up to about 250 GeV for $\Delta$M$<$5~GeV, which can neither be covered by LUX nor by XENON1T DM DD experiments. At the same time the XENON1T experiment will be able to complementarily cover the $\Delta$M$>$5~GeV parameter space up to large values of the DM mass, well beyond the LHC reach via monojet analyses.

\section*{Acknowledgements}
The work of VS and AB is supported by the Science Technology and Facilities Council (STFC) under grant number ST/L000504/1   and ST/L000296/1
respectively.
WP is supported by the Bundesministerium f\"ur Bildung und Forschung (BMBF)
under contract no. 05H12WWE.
 AKMB thanks Andreas Goudelis for helpful discussions.
 AB and DB thank Marc Thomas for useful discussions and help with solving of jet matching problems.